\documentclass[12pt,preprint]{aastex}
\begin{document}

\title{Photometric Variability in \textit{Kepler} Target Stars. III. Comparison with the Sun on Different Timescales.. }

\author{
Gibor Basri\altaffilmark{1},
Lucianne M. Walkowicz\altaffilmark{2},
Ansgar Reiners\altaffilmark{3}
}

\altaffiltext{1}{Astronomy Department, University of California, Hearst Field Annex, Berkeley, CA 94720}
\altaffiltext{2}{Department of Astrophysical Sciences, Princeton University, Peyton Hall, 4 Ivy Lane, Princeton NJ 08534}
\altaffiltext{3}{Georg-August-University Goettingen, Institute for Astrophysics, Friedrich-Hund-Platz 1, DE D-37077, Goettingen, Germany}
\begin{abstract}

We utilize \textit{Kepler} data to study the precision differential photometric variability of solar-type and cooler stars at different timescales, ranging from half an hour to 3 months. This is done in part by using the overall range of variability on long timescales of the light curves as an activity diagnostic.  We also define a diagnostic that characterizes the median differential intensity change between data bins of a given timescale. We apply these same diagnostics to solar data from SOHO that has been rendered comparable to \textit{Kepler}. In order to make a direct comparison between the Sun and \textit{Kepler} stars of different brightnesses, we develop a simple three-parameter noise model for \textit{Kepler} that is well matched to the lower envelope of variability for all stars and timescales in the \textit{Kepler} dataset. 

We come to the clear conclusion that the Sun exhibits similar photometric variability on all timescales when compared to similar solar-type stars in the  \textit{Kepler} field. We are also able to address again the question of what fraction of comparable stars in the  \textit{Kepler} field are more active than the Sun, and confirm that it is between a quarter and a third of them (depending on the timescale). The exact active fraction depends in part on what is meant by ``more active than the Sun'' and in part on the magnitude limit of the sample of stars considered. We argue that a reliable result can only be found for timescales of half a day or longer, and requires stars brighter than $M_{Kep}$ of 14, since otherwise there is too large a contribution of non-stellar noise.  We also confirm that as one moves to cooler stars, the active fraction of stars becomes steadily larger (greater than 90\% for the M dwarfs). We discuss the properties of different effective temperature groups of main sequence stars, from 6500-3500K. The Sun is a good photometric model at all timescales for those cooler stars that have long-term (monthly) variability within the span of solar variability.

\end{abstract}


\section{Introduction}
\label{sec:intro}

The \textit{Kepler} mission has proven its ability to open a new parameter space for the study of stellar magnetic activity. This is achieved through unprecedented precision in broadband photometry, coupled with extraordinary coverage in  length, cadence and continuity that had only previously been possible for the Sun. Early results \citep{Basri2010} for Quarter 1 indicated that on timescales of a month the Sun is a ``normal" star, in the sense that the bulk of solar-type stars observed by \textit{Kepler} fit within the variability range defined by the quietest to most active solar data. This variability range is defined simply as the span of differential intensities between the 5th and 95th percentile of the full range in a specified observing period. That early study used a simple polynomial technique to remove secular instrumental effects that were not well treated by the initial reduction pipeline. Although this treatment was relatively crude, it dealt with the major systematic effects and allowed a first look at the most important aspects of variability in the  \textit{Kepler} dataset. A subsequent study by \citet{Basri2011} examined the fraction of periodicity among the solar-type stars (limited to periods less than 2 weeks). This periodicity is often (but not always) related to rotational modulation due to starspots, and is the first step towards understanding the rotational periods of the tens of thousands of solar-types stars in the primary  \textit{Kepler} target list. A study in the same spirit has recently appeared for the COROT dataset \citep{Affer2012}. These studies find that many of the stars are periodic.

A subsequent study of \textit{Kepler} data by \citet{McQuillan2011} used a much more sophisticated correction to the raw data and confirmed the basic results of our two papers cited above. They suggest that the fraction of \textit{Kepler} stars that are more active than the Sun is larger than found by \citet{Basri2010}. We will discuss this discrepancy in Section \ref{sec:compsun}. A different sort of analysis by \citet{Gilliland2011} suggested in the same vein that the Sun is somewhat quiet compared to the \textit{Kepler} sample on timescales of a few hours (those most relevant to the search for planetary transits). There have been previous suggestions that the Sun might be photometrically quieter than the bulk of similar stars (eg. \citet{Radick1998}, although they were tentative.

Now the  \textit{Kepler} team has improved the pipeline reduction software substantially, so that for the most part stellar variability is preserved while instrumental effects are mitigated \cite{Stumpe2012}. The initial application of this was done on Quarter 9, which is also 3 times longer than the Quarter 1 dataset we examined previously. We now revisit how solar variability compares with the large sample of solar-type stars observed by  \textit{Kepler}, and do so for a variety of timescales. We would like to answer two basic questions: 1) in what sense is the Sun an average star when considering precision photometric variability, and 2) what fraction of the variability measured by \textit{Kepler} is intrinsically stellar, on different timescales and for stars of different brightness? Related to the first question, we also study the distribution of variability on different timescales to determine what fraction of the stars display a given level of variability. We look at stars with effective temperatures of 3500-6500K, and gravities that indicate they are main sequence stars based on KIC parameters \citep{Brown2011}. In order to best address the second question, we concentrate on the brighter end of the \textit{Kepler} sample, where stellar variability has a stronger contribution compared with other noise sources. 

\section{Method of Analysis}
\label{sec:method}
The fundamental dataset that we use is the  \textit{Kepler} Quarter 9 sample of exoplanet targets, reduced with the new PDC-MAP pipeline \citep{Smith2012, Stumpe2012}. This version of the light curve data has a number of instrumental effects removed or mitigated, including cosmic rays, discontinuities due to sudden changes in pixel sensitivity, secular drifts due to spacecraft motions which affect the distribution of light in the target aperture (which can be seen in stars geometrically related on the focal plane), and other electronic effects. A much more effective attempt has been made to preserve stellar variability. This can be seen by comparing the new pipeline data to the raw data, which easily showed the problems in the previous pipeline, as noted by \citet{Basri2011} and \citet{McQuillan2011}. We divide each light curve by its median then subtract unity to produce a differential intensity light curve. We then multiply by 1000 so that the units of differential intensity are parts per thousand (ppt); this puts them approximately onto a scale of millimags. We do this for subsets of the primary exoplanet target set which lie in the desired ranges of KIC temperature and gravity. The gravities are required to have log(g)$\ge$4.3, and the temperature ranges are given in Table 1. We further extract samples with given limits in \textit{Kepler} magnitude $M_{Kep}$ (designed to produce a sample of the brightest 1000 or so targets for each set of stellar parameters).  

\begin{figure}
\begin{center}
%\includegraphics[width=1.0\textwidth]{Virgo-Range.pdf} \\
%\subfloat[][]{\includegraphics[width=0.8\textwidth]{F1a.eps}} \\
%\subfloat[][]{\includegraphics[width=0.8\textwidth]{F1b.eps}} \\
\includegraphics[width=0.8\textwidth]{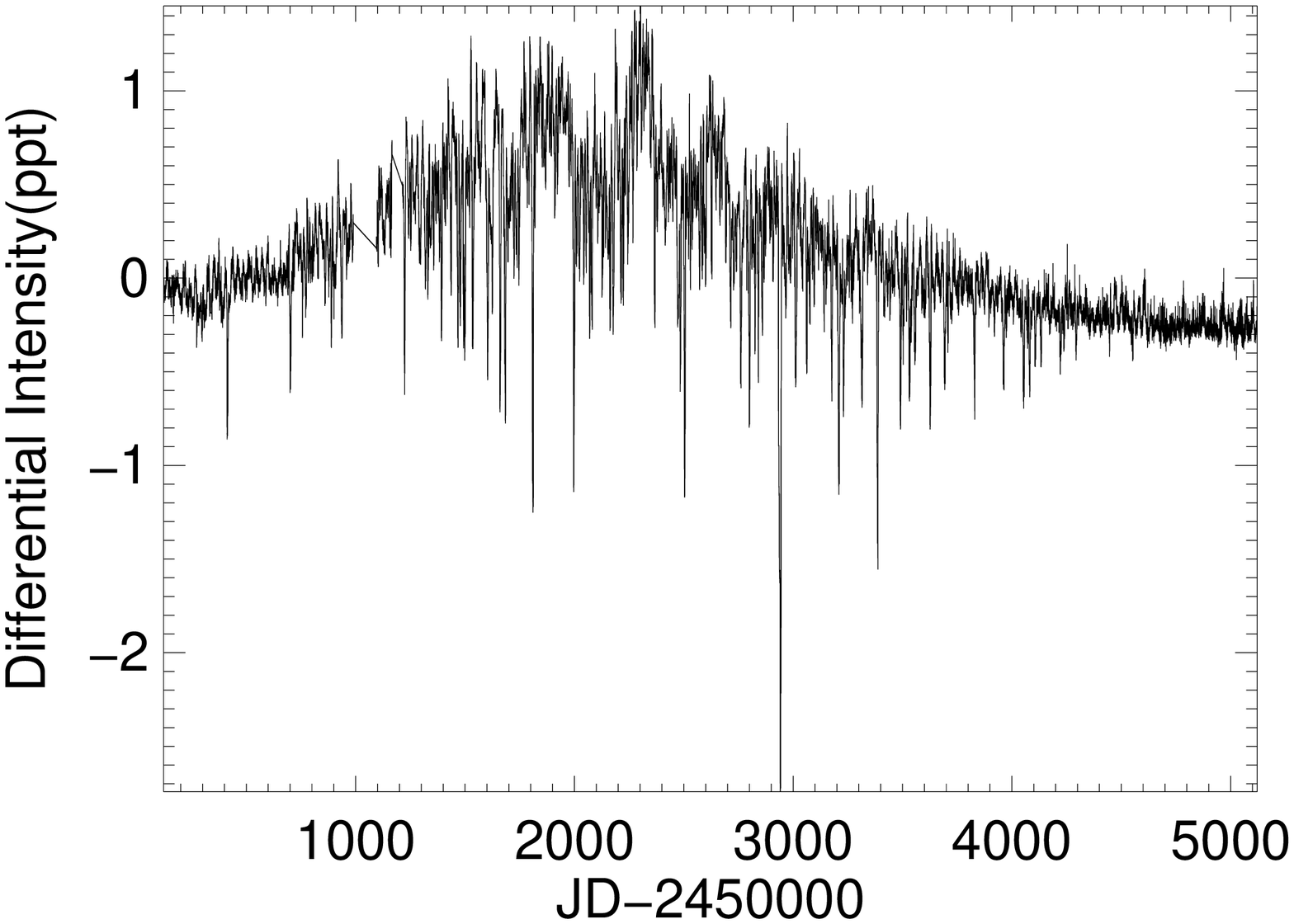} \\
\includegraphics[width=0.8\textwidth]{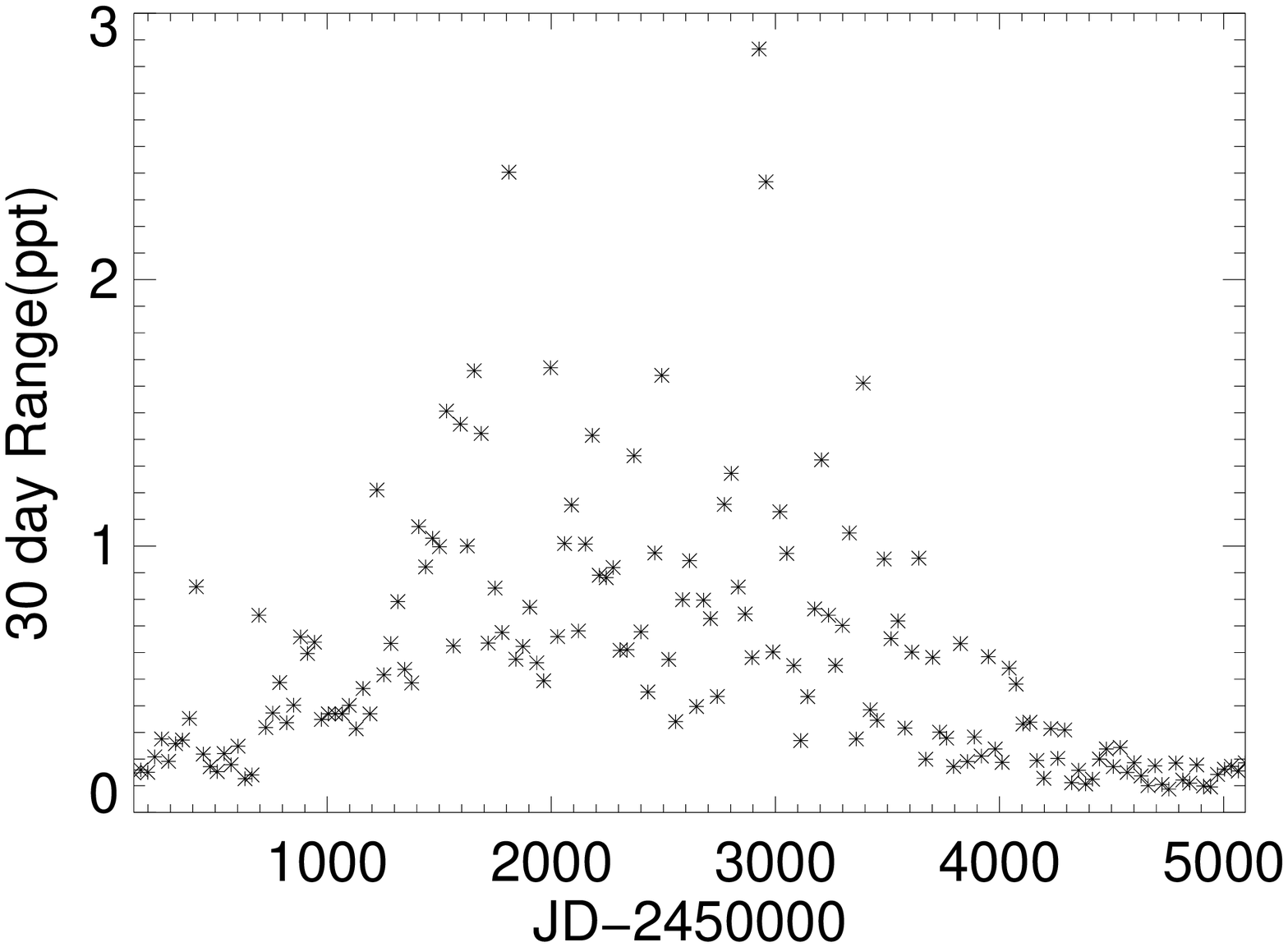} \\
\end{center}
\caption{The differential intensity variations for the Sun over Cycle 23 are shown in the upper panel. The units in both panels are parts per thousand. The lower panel shows the values for $R_{var}$ on a one month timescale for these data. }
\label{virgo}
\end{figure}

To compare the \textit{Kepler} sample with the Sun, we created a Kepler-like solar light curve using the Virgo data from the SOHO spacecraft. These data were obtained from from the ftp server ftp.pmodwrc.ch in the directory 
/data/irradiance/virgo/TSI/virgo\_tsi\_d\_ and\_virgo\_tsi\_h\_6\_002\_1204.dat. This data is a composite of data from the DIARAD and PMO6V absolute radiometers, calibrated to ``Level 2" \citep{Frohlich2001}. Its bolometric character is comparable to the  \textit{Kepler} data, which has a very broad visible bandpass (423-897 nm). One might worry that the UV contribution to the bolometric luminosity could introduce extra variability, but we found little evidence for a significant effect when comparing this to the data from the narrow-band Virgo photometers (see below). We linearly interpolated the hourly data into half-hour bins (to match the \textit{Kepler} cadence). This data stretches over all of Cycle 23 (1996-2009); we used it starting from JD2450121.5 for a length of 5000 days. The solar data was converted to differential intensity (ppt) in exactly the same way as \textit{Kepler} light curves; we did not normalize out the changes in irradiance between quiet and active phases (the median used to normalize was for the entire cycle). The solar differential light curve can be seen in the upper panel of Fig. \ref{virgo}.

Previous  \textit{Kepler} analyses have used Virgo SPM data. This data is taken in two filters (not using the ``blue'' filter) with bandwidths of 5 nm centered at 500 and 862 nm. That is in stark contrast to the broad \textit{Kepler} bandpass. We argued \citep{Basri2010} that nonetheless the data are comparable, since to first order the irradiance variations look like slight changes in effective temperature (and the amplitudes in the two filters are related by the inferred $\Delta$T at the appropriate wavelengths in the Planck function). We take a second look at this question here. Converting the changes in flux with temperature to changes in counts we find again that a sum of the two filters is a reasonable approximation to what one gets by integrating over a large bandpass using a photon-counting (not bolometric) detector. The extra sensitivity of the shorter wavelength to temperature is mostly offset by the larger number of counts at longer wavelengths. 

We can, for example, mimic what is done when using the two VIRGO filters, by summing the counts per second for a blackbody of solar effective temperature at their two wavelengths, then calculating the difference in that sum after changing the temperature by $\Delta$T (eg. adding 1 degree Kelvin). We use the difference because the astronomical measurements we are using are differential intensities. To mimic what is done by the  \textit{Kepler} photometer, we can instead sum the mean counts for the solar temperature at 400 wavelengths evenly spaced between the two filter wavelengths, and then take the difference with the sum obtained after adding $\Delta$T. The ratio between these two differences is within one percent of unity, which implies that one makes a differential error of only one percent when using the two wavelengths compared with using 400 wavelengths. We are therefore confident that simply adding the SOHO data for the two filters and converting to a differential intensity is a reasonable translation from SOHO to Kepler data. This is borne out by a direct comparison of overlapping re-binned versions of SPM and DIARAD SOHO data (DIARAD is a very broad-band bolometer). Thus the solar data in this paper are compatible with the previous comparisons between \textit{Kepler} and the Sun. 

 One point of difference between the data in Figure \ref{virgo} and that in \citet{McQuillan2011} is that we have normalized to the median of the whole dataset (preserving the overall change in irradiance between quiet and active Sun), whereas they removed the basic trend in irradiance between the quiet and active Sun. That is, we used a single median for normalization while they used a running median to flatten the dataset on timescales long compared to the rotation period (they also only utilized data for the first 2200 days or so). We fit a 7th order polynomial (which only affects timescales well longer than a rotation period) to our full dataset and subtracted it to reproduce the effect of the running median (thus producing a similarly flattened differential light curve). We found that $R_{var}$ (calculated on the rotational timescale of one month) was not significantly affected by the flattening procedure (meaning that the changes were small compared to the point to point variations of $R_{var}$), so it doesn't really matter for purposes of comparison with \citet{McQuillan2011}.

\subsection{Measuring Variability}
\label{sec:varmes}
The measure of variability that has been used since the development phase by the  \textit{Kepler} project is called CDPP (combined differential photometric precision). It is used to provide a composite measure of sources of noise in the photometric signal (where stellar variability is considered one such source) with an eye to estimating the probability that a planetary transit can be detected. Very detailed calculations and engineering data go into the assessment of the contributors to CDPP \citep{Christiansen2012}.  As they describe, it includes harmonic filtering in frequency space (one purpose of which is dampening stellar signals), time-variable signal-whitening using wavelets, and a sophisticated methodology whose overall effects are hard to disentangle. It is not the ideal measure of variability if one is primarily interested in stars, because of its complicated nature and the difficulty of assessing the ways in which the stellar variability (which will have different timescales and shapes for different stars) interacts with other noise sources and is altered by the methodology. In \citet{Gilliland2011} CDPP was approximated as something like the standard deviation of intensity within blocks of data on a given timescale, after the blocks have had a 2-day running polynomial around each point subtracted off (and very deviant points are eliminated); see that paper for a fuller explanation. The Savitsky-Golay filter reduces stellar variability on timescales of a day or two, but not completely and in a star-dependent way. The purpose of calculating CDPP is to estimate the level of noise that interferes with transit detection; it was not designed to study stellar variability. In particular the value of CDPP depends on the behavior of the star in frequency space, which is then subjected to manipulations in a manner that is difficult to connect to the star's actual behavior on a given timescale.

We employ two very simple basic measures of variability in this analysis. They are intended to be direct measures of changes in differential intensity on specified timescales (and so are computed in the time domain). We are interested in stellar activity, so our diagnostics connect fairly directly to physical changes on the star that are not strictly periodic, such as rotation or evolution of starspots, flares, and the like. A separate approach is employed by others to study stellar pulsations and asteroseismology (which are better suited to frequency space). The first of our diagnostics has been defined by \citet{Basri2011} as the ``range'' $R_{var}(t_{len})$: a measurement which assesses the span of differential photometric changes in a light curve over a given length in time. It is found by sorting all the differential intensity points for the light curve and taking the difference between the 5\% and 95\% values (to avoid anomalous drops or peaks). We have found that this measure can be considered a metric of stellar photometric activity; more spotted stars will have higher $R_{var}$ and a larger fraction of stars with high $R_{var}$ are periodic and have shorter periods. Of course, $R_{var}$ can also be large if there are frequent or lengthy enough eclipses (transits will only affect $R_{var}$ if they are numerous) or if the star is a pulsator or red giant. It is worth noting, in addition, that $R_{var}$ can also be affected by instrumental noise when that noise is comparable to or larger than stellar variability (this becomes increasingly true for fainter stars), or if secular drifts are too large over the chosen timescale. The $R_{var}$ for the Sun on a one-month timescale is shown in the bottom panel of Fig. \ref{virgo}. The behavior of $R_{var}$ tracks quite closely with other measures of whether the Sun is ``active" or ``quiet", and the solar cycle is very obvious. Thus, for the purposes of this paper, we are {\it defining} ``activity" to be measured by $R_{var}$(30 days); a month is close to the solar rotation period. For the \textit{Kepler} stars, we have 3 months of data. We computed $R_{var}$ for each one month segment and used the median value if we needed a single value per star (but used all the values when looking at histograms).

Our other primary measure of variability is the median differential variability MDV($t_{bin},\Delta t$). The MDV is a very simple measure of the bin-to-bin variability for bins of a given timescale $t_{bin}$. One could shift the comparison between bins by $\Delta t$ (a bit analogous to the galaxy spatial correlation function, but temporal); here we will confine ourselves to $\Delta t = t_{bin}$. Thus MDV is obtained by re-binning the light curve onto the desired timescale ($t_{bin}$), then compiling the absolute difference between all adjacent bins. Each bin contains the average (mean) of the points that go into it. We define MDV($t_{bin}$) as the median value of those absolute differences. For a light curve of pure uniform noise, MDV($t_{bin}$) scales as $1 \over \sqrt(t_{bin})$, much like Poisson noise with different integration lengths. The dispersion of realizations of MDV($t_{bin}$) along a light curve grows with $t_{bin}$, however, since there are fewer samples for large $t_{bin}$ in a given light curve. For example, for 100 light curves 90 days long with half-hour sampling that each consist of uniform noise with a standard deviation of unity, the standard deviation in MDV for all the light curves is about 0.02 independent of $t_{bin}$, but the value of MDV(0.5 hr) (point-to-point) is 0.95 and the value of MDV(8 days) falls to 0.04. Thus the standard deviation in the value of MDV(0.5 hrs) in this example is about 2\%, but it is 50\% for MDV(8 days).

The MDV is a combined measure of both instrumental noise (meaning all non-stellar sources of variability, including photon noise due to collecting area) and any contribution of stellar variability. It will be more sensitive to instrumental noise for shorter timescales or for low levels of stellar variability; for longer timescales we are essentially integrating starlight longer and that increases the S/N. The value of the MDV at a given timescale also depends on the timescales for stellar variability. To appreciate the difference between MDV and $R_{var}$, imagine a star whose brightness steadily increases. Over a month the star might accumulate a large $R_{var}$, but the MDV would depend linearly on the timescale used to evaluate it. Now suppose the star has the same MDV on the chosen timescale, but the derivative switches between positive and negative on that timescale. The $R_{var}$ would become much smaller while MDV is unchanged.  On the other hand, if the star is nearly constant on the shorter timescales but more variable on the longer timescales, the MDV could increase with timescale (even though the S/N is better). 

In the case where photometric variability is largely due to the passage of starspots on the visible hemisphere due to stellar rotation, the MDV will depend partly on the timescale of rotation and partly on the amplitude of variability (set by both spot coverage and the extent to which the spot distribution is asymmetric). For timescales quite short compared to a rotation the MDV will tend to depend more on the amplitude (which is related to $R_{var}$); when the timescale becomes comparable to the rotation period the relation between MDV and $R_{var}$ can be more of a mixture of spot pattern and spot amplitude. We will use MDV to compare the \textit{Kepler} targets to the Sun; this will elucidate whether they are different than solar in this very straightforward method of comparison on each timescale. Our other purpose is to tease out the contribution of instrumental noise compared with stellar variability. A powerful signature of non-stellar noise is whether the MDV is a function of apparent magnitude on a given timescale. To the extent that it is, that is a strong sign of a non-stellar contribution, since stars know nothing about their apparent magnitude. Implicit in this last statement is the presumption that we are considering stars with similar intrinsic luminosities and variability distributions, which we do by taking restricted temperature bins of main sequence stars.

\begin{figure}
\begin{center}
\includegraphics[width=1.0\textwidth]{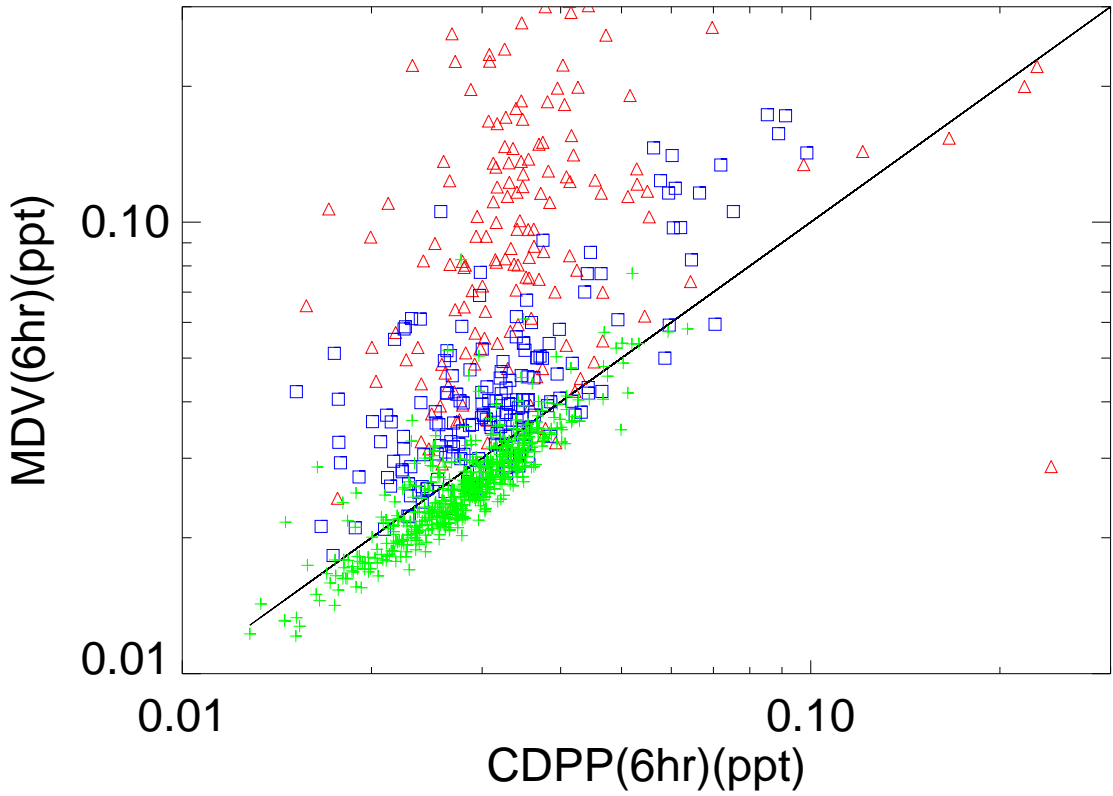} \\
\end{center}
\caption{A comparison of CDPP and median differential variability (MDV) at 6 hrs of a bright G dwarf \textit{Kepler} sample. The line is the locus of equality. The symbols are coded by whether they have a one-month $R_{var}$ less than the active Sun (green plusses), in the range of the active Sun (blue squares), or greater than the active Sun (red triangles). }
\label{cdppmdv}
\end{figure}

In Fig. \ref{cdppmdv} we show a direct comparison between the pipeline CDPP and MDV at a 6 hour timescale (\citet{Gilliland2011} actually used 6.5 hours, but this difference is not important). The \textit{Kepler} points are for a sample of bright G dwarfs (described in the next section). They have been sorted (somewhat arbitrarily) by $R_{var}$: the green plusses (quiet) have $R_{var}$(30 days) less than a minimum active SOHO value (0.8 ppt), the red triangles (very active) have $R_{var}$ greater than a maximum SOHO value (1.7 ppt), and the intermediate blue squares are within the range of the active SOHO points (see Fig. \ref{virgo}). In comparing MDV to CDPP there are a couple of salient points. First, the MDV is a great deal better at sorting out which stars are more active (in the sense of having a high $R_{var}$) than CDPP. There is a slight separation between the least active and most active stars in CDPP, but a great deal of mixing. This is inherent in the design of CDPP, but it means that CDPP is not really capable of sorting out the contribution of stellar variability to the derived value (stars with rather different levels of variability have the same value of CDPP). Second, the quiet stars have MDVs which track CDPP (at values that are a little lower on the same differential intensity scale, so there is an inherent normalization difference). Intermediate and high activity stars have MDVs which are greater than their CDPP (by amounts which depend on the stars' $R_{var}$, and the correlation weakens as stellar variability gets higher). One must therefore be rather careful in making a direct comparison between conclusions based on each of these two variability measures. Our claim is that if one is interested in assessing stellar variability (or the contribution of stars to observed variability) then MDV is a more sensitive and appropriate diagnostic. In particular, we believe that this difference in the sensitivity of the diagnostics is one of the primary reasons that we reach different conclusions than \citet{Gilliland2011} regarding whether the Sun is different from the stars in activity level. Our conclusions do not change if we use the version of CDPP in that paper, or the values supplied by the pipeline. The distribution of stars in CDPP is simply not an accurate reflection of their astrophysical variability distribution (as assessed in much more straightforward ways).

\begin{figure}
\begin{center}
\includegraphics[width=1.0\textwidth,height=7.0 in]{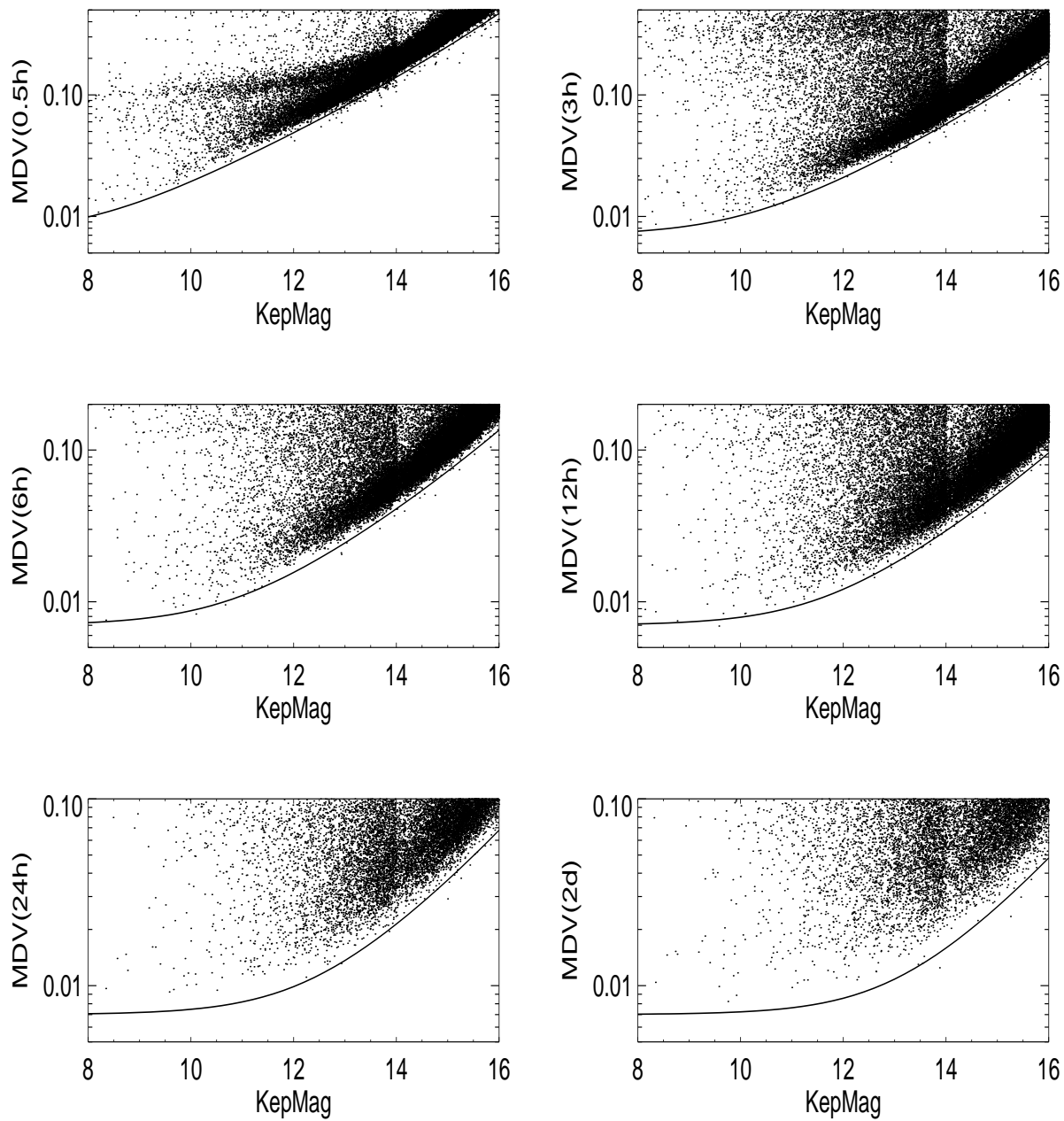} \\
\end{center}
\caption{The median differential variability of the full \textit{Kepler} sample in Quarter 9 for a variety of timescales. The solid line in each plot is the fit of our noise model to the bottom envelope of the data. Units for MDV are always parts per thousand (ppt). Only every third point is plotted, which makes the sparse set of very low points look even sparser. }
\label{noisemod}
\end{figure}

\subsection{A Noise Model}
\label{sec:noisemod}
Our intention here is to be as observationally empirical as possible (removing the need for engineering data). This is in the spirit of the approach of \citet{McQuillan2011}. We sometimes call all noise not arising intrinsically from the stars ``instrumental";  the largest component of this non-stellar noise is the easily quantified Poisson statistics contribution from photon noise. We presume a simple model for sources of variability: there is a noise source which depends on stellar brightness like the square root of the signal, there is a noise source that is independent of brightness, and there is a term due to stellar variability (also independent of brightness). We calibrate the amounts of each by considering the entire  \textit{Kepler} sample for each MDV($t_{bin}$), and fitting to the envelope of least variable points at all magnitudes (which we presume express the minimum stellar variability along with noise). The equation for MDV($t_{bin}$) is 

\begin{equation}
MDV(t_{bin})={\{ V_*(t_{bin})^2+ 0.95^2 ( {{\sqrt{\delta I(t_{bin})}+\sqrt{N_0(t_{bin})} }\over \delta I(t_{bin}) } )^2 }\} ^{1\over 2}
\end{equation}

The factor of 0.95 is that found in the previous subsection for uniform noise, and the division by the differential intensity $\delta I(t_{bin})$ is needed because the light curves are normalized by their medians prior to computing differential intensities. $N_0$ is the brightness-independent instrumental noise term, and $V_*(t)$ is the contribution of stellar variability, which we take to be composed of $V_*^{min}+V_*^{act}(t)$. $\delta I$ depends on \textit{Kepler} magnitude $M_{Kep}$  (magnitudes in the broadband \textit{Kepler} images) like

\begin{equation}
\delta I(t_{bin})=\delta I_{0}(t_{bin}) \times 10^{({12-M_{Kep}}) \div 2.5} 
\end{equation}

where $\delta I_0(t_{bin})$ is the minimum differential intensity term (due primarily to photon noise) for $Kepler$ magnitude of 12 (as seen in the whole sample) for a given $t_{bin}$. We found that  $\delta I_0(t_{bin})$ could be defined for the basic long cadence ($t_{bin}$=0.5\rm{hr}) and then $\delta I_0(t_{bin})=\delta I_0(0.5\rm{hr})*({t_{bin} \div 0.5\rm{hr}})$. The same is true for $N_0(t_{bin})$ (in the same units as $\delta I_0(t_{bin})$). Thus, our noise model is defined by a stellar contribution and only the two free parameters  $\delta I_0(0.5\rm{hr})$ and $N_0(0.5\rm{hr})$. 

In order to determine these two free parameters, we looked at the MDV($t_{bin}$) of the full \textit{Kepler} sample for each $t_{bin}$ of interest. The values of $t_{bin}$ start with the long cadence itself (0.5 hours), and the next value is 3 hours. After that we take each value of $t_{bin}$ and multiply by 2 to get the next value, up to a maximum of 8 days. In Fig. \ref{noisemod} we show these datasets up to 2 days. We adjust the values of $\delta I_0(0.5\rm{hr})$ and $N_0(0.5\rm{hr})$ to match the bottom envelope of the data points. The bottom is defined by taking normalized histograms of MDV values in 0.4 mag bins and finding the location in log(MDV) where the lowest 1\% of points are bounded below. A third-order magnitude-dependent polynomial is then fit to these points.The level and basic shape of the curve is controlled primarily by $\delta I_0(t)$, while $N_0(t)$ mostly affects its curvature (through the changing ratio between instrumental and photon noise). The presumption  is that the stellar contribution is minimal for this bottom envelope, and that except for the bright stars the value of the lowest points is indicative of the instrumental noise rather than intrinsic stellar variability. If that were not true the bottom envelope would not steadily rise to fainter magnitudes; intrinsic stellar variability is flat across apparent magnitudes. An exception is the effect that cooler stars are more present at faint magnitudes and more variable. We test this later by looking at different temperature bins. 

At 2 days or longer the stellar variability seems to be dominant above 14th magnitude and the bottom envelope begins to lift off the fit found for shorter bins. Using 12th magnitude as a fiducial the free parameter values are $\delta I_0(0.5\rm{hr})=5.\times10^8$ and $N_0(0.5\rm{hr})=1.\times 10^7$ (before conversion to ppt). We first tried finding best fit values for these parameters at each timescale separately. It became clear, however, that it was reasonable to use a single pair of values (the variation of these values for individual fits was less than 20\% over all the timescales, and the fits they produced were not noticeably better). A good fit is obtained for all values of $t_{bin}$ with these values (along with a single value for $V_*^{min}$). This implies that we have captured the essence of the behavior of the instrument as manifested in differential intensity changes in long cadence data on the various timescales. By construction we have the right noise level at all the magnitudes considered. We can compare these values to those expected from the Kepler instrument as given in \citet{Gilliland2011}. Their value for the variance due to photons is $4.5\times 10^9$ in 6.5 hours, with a readout noise term about ten times smaller. Our value for 6.5 hours is $6.5\times 10^9$ which is roughly consistent; of course their noise model is much more complex so there is no reason why our proxies for photon counts or noise should agree exactly. 

For $V_*^{min}$ we find a best fit value for the 3 and 6 hour cadences, using the trends between 8th and 10th $M_{Kep}$, which show a flattening at these bright magnitudes. This flattening is presumably due to the dominance of stellar variability over instrumental effects at these bright magnitudes; such a flattening does not quite occur for the half hour cadence (which presumably is not quite a long enough integration to allow the quietest stellar variability to dominate at these magnitudes). A single value of $V_*^{min}=.01$ (or 10 ppm) was adequate for all the time bins . That did not necessarily have to be the case, since the variability could have different characteristics on different timescales (which it does for $V_*^{act}(t_{bin})$). In fact $V_*^{min}$ is not well-determined at all for the longer timescales, since there are very few stars along the bottom boundary (stellar activity begins to dominate for those timescales); a value of 0.005 is perhaps slightly superior for the several day timescales. 

\section{Comparison of the Sun to the \textit{Kepler} Solar-type Sample}
\label{sec:compsun}

We would now like to compare a sample of \textit{Kepler} stars with the Sun, to see whether the Sun exhibits ``normal'' photometric variability compared to main sequence stars similar in temperature in the  \textit{Kepler} field. For this purpose we prefer a bright sample (\textit{Kepler} magnitudes $M_{Kep}$ no fainter than 12.5) to minimize the photon noise. We choose stars out of the planetary search targets that have KIC parameters which are solar-type:  log(g) no smaller than 4.3, temperatures between 5500-6000K. This yields a sample of 974 targets in Q9. We computed the MDV($t_{bin}$) for all of them at all the chosen timescales, and computed the same quantities for our samples of SOHO data described in Section \ref{sec:method}. 

\begin{figure}
\begin{center}
%\includegraphics[width=1.0\textwidth]{GdwfVar(t)-Range.pdf} \\
%\subfloat[][]{\includegraphics[width=0.8\textwidth]{F4a.eps}} \\
%\subfloat[][]{\includegraphics[width=0.8\textwidth]{F4b.eps}} \\
\includegraphics[width=0.8\textwidth]{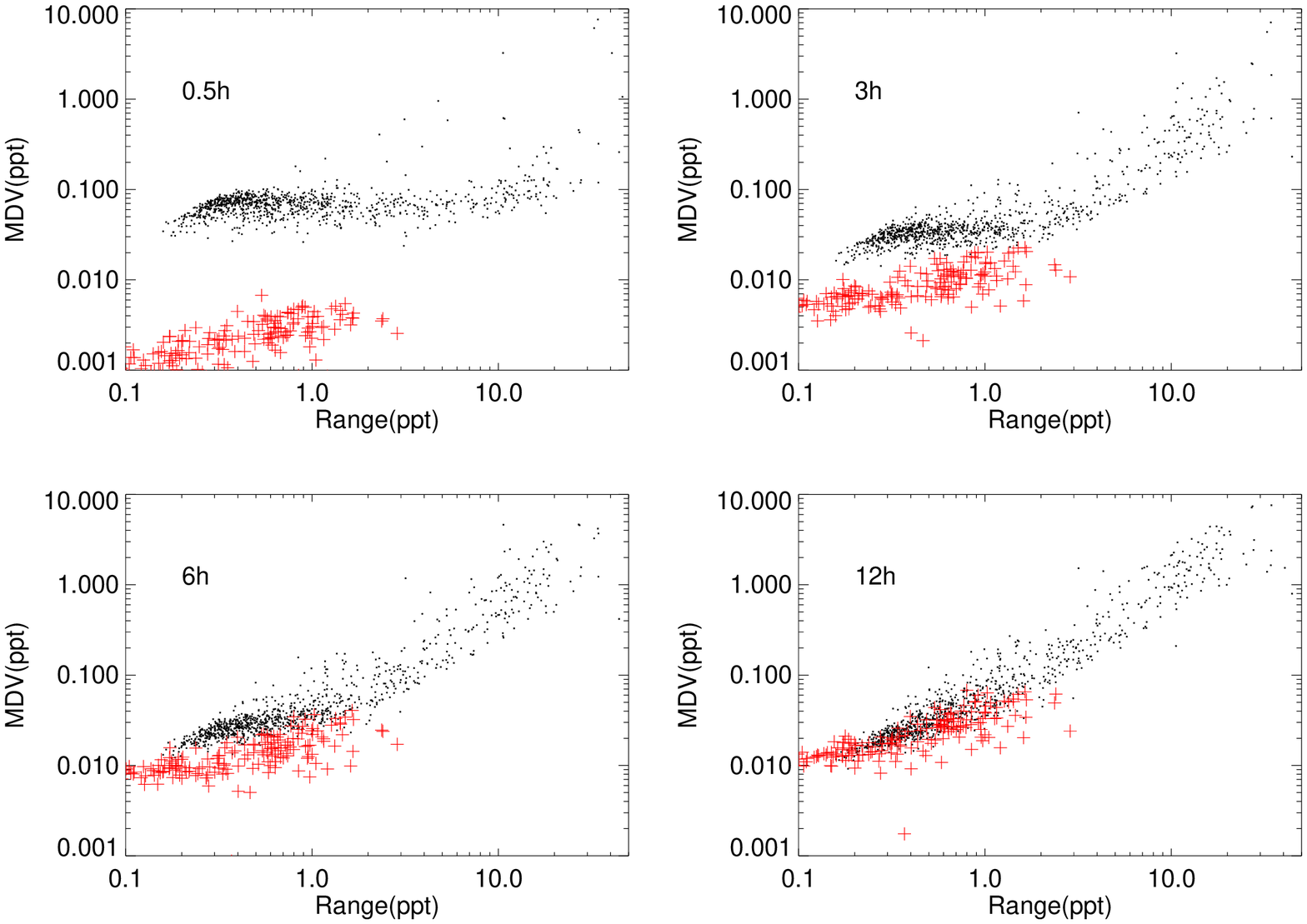} \\
\includegraphics[width=0.8\textwidth]{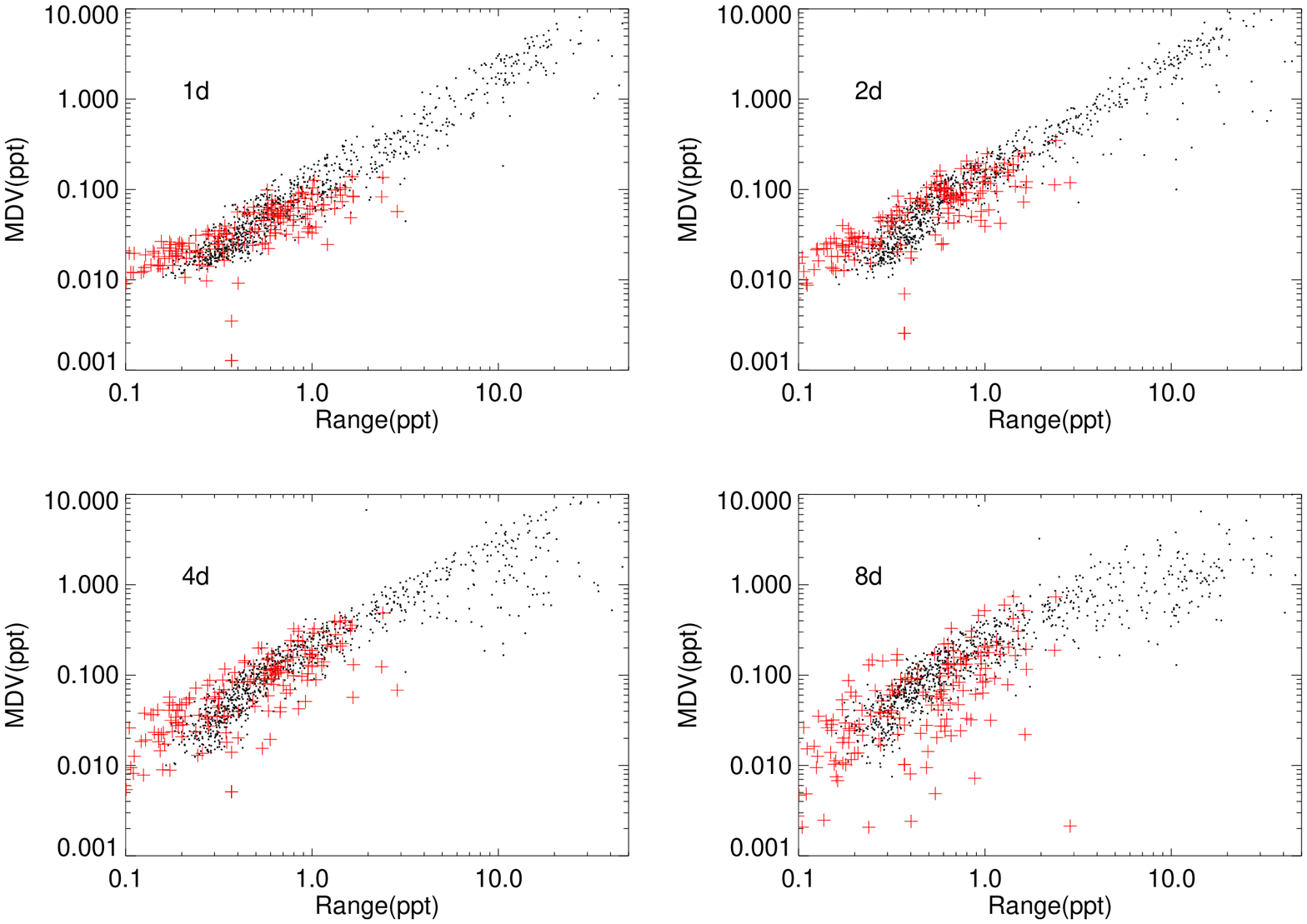} \\
\end{center}
\caption{The variability on different timescales of G dwarfs as a function of $R_{var}$ (amplitude of variability over 30 days). The points are \textit{Kepler} stars and the red plusses are SOHO data for the Sun over a whole cycle in 30 day increments.  }
\label{rangemdv}
\end{figure}

In order to compare the Sun to \textit{Kepler} data, we need some way of estimating the likely MDV($t_{bin}$) that the Sun would display if observed by Kepler at different brightnesses. In Fig. \ref{rangemdv} we show the bright G-dwarf sample plotted with $R_{var}$ on the abscissa and various values of MDV($t_{bin}$) on the ordinate. The SOHO points are shown in red, using the same variables and units. It is beyond the scope of this paper to directly compute the expected instrument noise for SOHO on different timescales, but it is clear from looking at the 0.5\rm{hr} plot that the SOHO data are an order of magnitude quieter than all the \textit{Kepler} points. This didn't have to be the case in principle, but is perhaps not surprising since SOHO is  looking at a MUCH brighter object, albeit through narrow band filters. There is no difference in MDV(0.5hr) between the quiet and active Sun. This is also not a surprise; that timescale is shorter than almost all integrated magnetic activity would produce (while magnetic activity is what is responsible for the spread in $R_{var}$). The \textit{Kepler} points  also show little percentage variation as a function $R_{var}$ (with a few exceptions). The explanation is presumably the same as for the Sun in the same span of $R_{var}$, but it extends to substantially greater values of $R_{var}$ as well. The gap between the solar and $Kepler$ points is easily explained if the smallest intrinsic stellar variations (which might share a floor similar to the Sun) are overwhelmed by noise. This becomes even more compelling an explanation as we look to longer timescales.

As we increase the timescale the photon noise is integrated down in importance compared with stellar variability (which becomes more influential at longer timescales). The \textit{Kepler} data is still dominated by noise (flat) for stars with solar $R_{var}$ for MDV(3hr) and no more variable than the active Sun, although the excess noise is now only a factor of 2-3. Since the binning is 6 times greater now, the expected decrease in noise is about 2.5. At the same time, the Sun shows a little more variability, presumably picking up a little greater contribution from magnetic activity. The very quietest solar values (with smallest $R_{var}$) appear just above 0.01 (and remain there for longer timescales). These are instances where there just isn't much magnetic activity present. The Sun begins to show a correlation between MDV(3hr) and $R_{var}$ in its more active phases. Stars more active than the active Sun do show a correlation between MDV($t_{bin}$) and $R_{var}$, presumably because for these very active stars the effects of rapid rotation and large spot coverage begin to affect the derivative of the light curve on this timescale. Moving to MDV(6hr), the Kepler points again drop by about the expected amount if caused by integrating noise. The average \textit{Kepler} solar-type star still exhibits a larger MDV than the average Sun at a comparable $R_{var}$ at transit timescales (6-6.5hr), in concert with the results of \citet{Gilliland2011} (although they were talking about CDPP). This appears to us to be an instrumental rather than stellar effect, given the gaps between the Sun and the stars at shorter timescales and their behavior, which is what is expected if  noise-dominated. We cannot think of a physical (stellar) explanation for the way this gap behaves at short timescales.

The quiet and active SOHO and \textit{Kepler} values at comparable $R_{var}$ come into good alignment by MDV(12hr) (giving us confidence that this methodology of comparison is valid). The correlation between $R_{var}$ and MDV($t_{bin}$) is fully established, suggesting that at this timescale the \textit{Kepler} data are reflecting primarily stellar (magnetically driven) variability. The lack of a gap between the Sun and the stars also suggests that now we are dominated by stellar effects. The other conclusion one can draw is that the variability on various timescales is correlated; a more ``active'' star is more active on most timescales (with some individual scatter). These data do not suggest that the Sun is unusually quiet compared to its neighbors for the same $R_{var}$, although the Sun extends to quieter $R_{var}$. The correlation between MDV($t_{bin}$) and $R_{var}$ is present at all timescales of half a day or longer. The Sun shows a greater range of MDVs at 4 and 8 days than the Kepler stars do (perhaps the $Kepler$ data are subject to an unknown systematic effect at these timescales). On timescales of a month some of the solar points are a little quieter (have smaller $R_{var}$) than any of the \textit{Kepler} stars. In these cases (from the recent extended and very quiet solar minimum) the Sun shows no obvious spot dips, but a very small rotational modulation is still perceptible (which probably would not be visible for stars given current systematic effects at long timescales in the \textit{Kepler} pipeline).

\begin{figure}
\begin{center}
\includegraphics[width=1.0\textwidth]{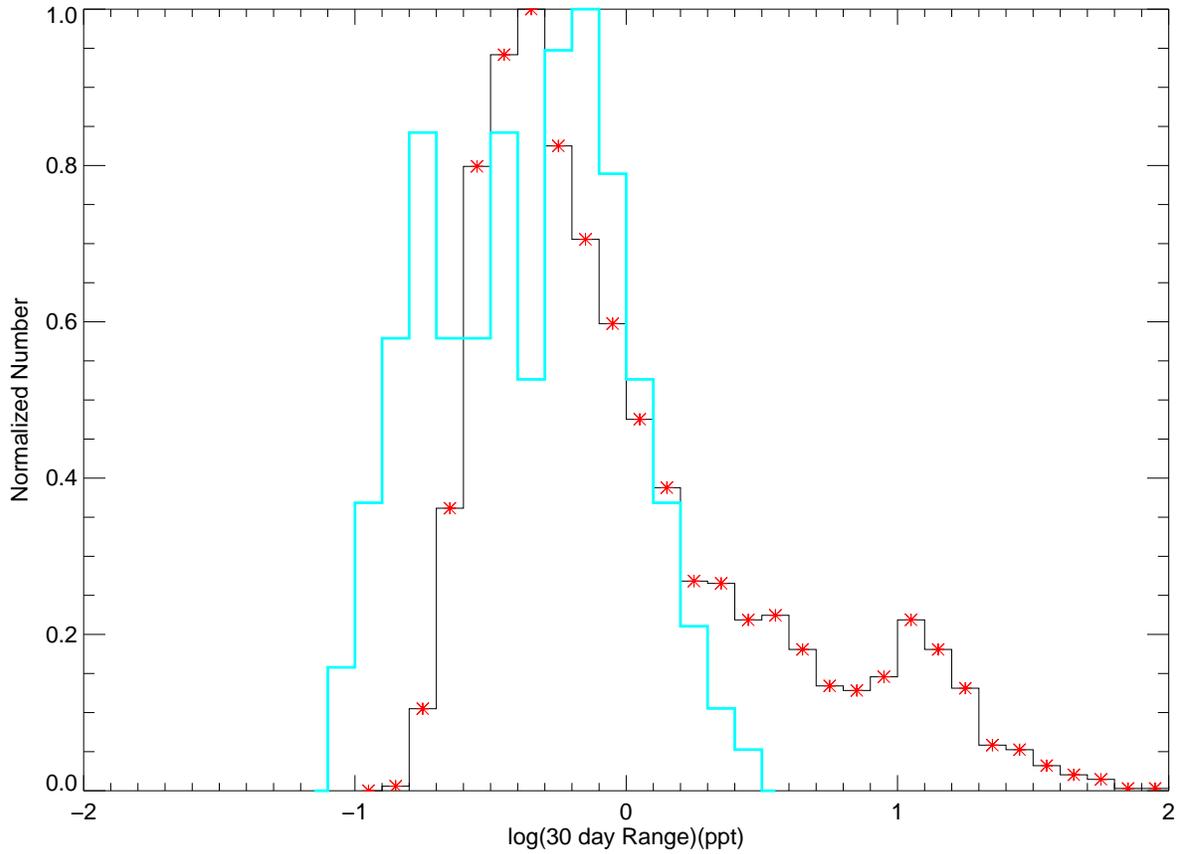} \\
\end{center}
\caption{The variability as measured by one-month $R_{var}$ for the Sun (thick line) and  \textit{Kepler} G dwarfs (starred). The histograms are normalized to the same peak values. They are comparable because the solar data provide snapshots at all cycle phases for the Sun, while the \textit{Kepler} data sample solar-type stars at all random phases of their cycles. The comparison provides a means of estimating what fraction of stars could be somewhere in a solar-like cycle, and what fraction are clearly more active or quiet. }
\label{histogram}
\end{figure}

\begin{deluxetable}{ccc}
\tablecaption{Fraction of Stars Around Various R$_{var}$(30d).  }  \tablenum{1}  \tablecolumns{3} \tablewidth{0pc}
\tabletypesize{\footnotesize}  
\label{perctab}
\tablehead{ \colhead{Solar \% \tablenotemark{a}} & \colhead{Stellar \% \tablenotemark{b}} & \colhead{log(R$_{var}$(30d)) (ppt)} { \tablenotemark{c}} }
\startdata  
(20) 80 & 96 & -0.62 \\
(30) 70 & 87 & -0.52 \\
(40) 60 & 60 & -0.31 \\
(50) 50 & 49 & -0.19 \\
(60) 40 & 45 & -0.14 \\
(70) 30 & 41.5 & -0.08 \\
(80) 20 & 36 & 0.00 \\
(90) 10 & 29 & 0.12 \\
(95) 05 & 25 & 0.23 \\
(100) 00 & 21 & 0.40 \\
\enddata
\tablenotetext{a}{The percent of solar points which lie (below) above the range in column 3. }
\tablenotetext{b}{The percent of stellar points which lie above the range in column 3. }
\tablenotetext{c}{The range at which the percentages in the other columns is computed. }
 \end{deluxetable}

\subsection{The Active Fraction of G Dwarfs}
\label{sec:actfrac}

A very illuminating use can be made of these data by computing histograms of the one-month $R_{var}$ for the Sun and stars. This is productive because the solar data are 161 samples of the behavior of the Sun (a month at a time) at all parts of its activity cycle. The \textit{Kepler} data, on the other hand, represent a large number of stars during the same month, which should also sample all phases of the ensemble of activity cycles. We normalize the histograms to the same peak value so that the different sample sizes don't affect the comparison too much (Fig. \ref{histogram}). It is gratifying to find that the basic shape of the histograms are similar. We can use this to look at the fraction of stellar points that lie outside the solar points (or vice versa). There is the aforementioned excess of very quiet solar points; this is not due to noise, which doesn't appreciably affect $R_{var}$ for the Sun or bright \textit{Kepler} stars.  Another feature of possible note is that the solar histogram has a dip at the range where the stellar histogram peaks. This is a range where the Sun is transitioning from quiet to active. 

As noted above, 20\% of solar points are quieter than any stellar points, and 20\% of stellar points are more active than any solar points.  This low noise tail arises almost exclusively in the most recent solar minimum (which was not included in the datasets used before this paper).We suspect the quiet end of the stellar data may still be subject to instrumental systematics. We have looked at the distributions at the other timescales (several are apparent in Fig. \ref{rangemdv}). At the shortest times, the Sun is altogether quieter due to the superior S/N of the SOHO data. By 12 hours there is only a slight quiet excess, and there is none again until 8 days. This might be an indication that the Kepler pipeline is still introducing a spurious low-level variance at timescales of a week or longer. If true it could explain the fact that 20\% of the solar points have a lower range (which is a long timescale diagnostic) than any of the Kepler points. 

It is clear that the Sun and the stars share a median range and most of the points in both sets lie at similar ranges. The actual values of the ranges for various percentages of the solar points, and the fraction of solar-type \textit{Kepler} stars that are above each of them are given in Table \ref{perctab}. There is an obvious tail of stars that lie at higher $R_{var}$ than any of the solar points. We believe this tail is astrophysical; these are the stars that are clearly more active than the Sun. We can use the point at which the solar distribution drops off rapidly to define the high limit of solar variability, and then ask what fraction of the \textit{Kepler} stars lie above that. This approach has the virtue that we have essentially accounted for stars that might be in low or high parts of their cycle by comparing them with the Sun at all parts of its cycle. We find that roughly that 30\% of stars are more active than 90\% of the observed solar ranges, and 20\% of stars are more active than 100\% of observed solar ranges. This is consistent with what \citet{Basri2010} found, but significantly lower than what \citet{Gilliland2011} or \citet{McQuillan2011} found. They both claimed that about 60\% of stars are more active than the Sun. We find instead that 60\% of the \textit{Kepler} stars are only more active (have larger one-month ranges) than about that same fraction of solar points. 

\begin{figure}
\begin{center}
\includegraphics[width=1.0\textwidth]{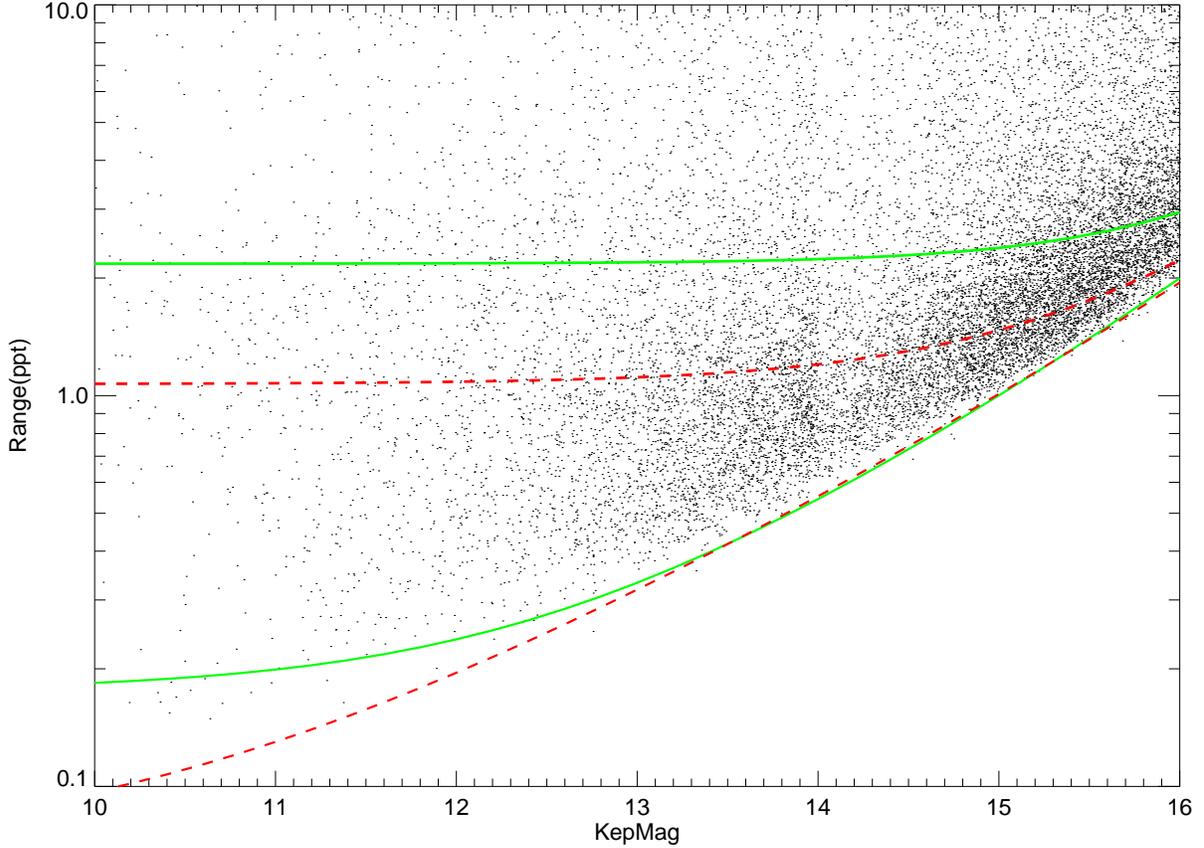} \\
\end{center}
\caption{A comparison of our results with that of McQuillan et al. (2011; their Fig. 4). Here we consider the full range of magnitudes for a sample G dwarfs, with the ordinate as R$_{var}$(30d). The lower solid line is our noise model and the lower dashed line is theirs (the main difference being our inclusion of a minimum level of stellar variability which only affects the bright end). The upper dashed line is the level of solar variability they take (based on the average active Sun over two years), while the upper solid line is our chosen estimate for the most active Sun. Rather different active fractions are found, mostly because of what happens at the faint end, where a lower estimate for the Sun causes faint stars to be squeezed above it by the increasing influence of noise. The fact that the solar lines curve up fainter than 14th magnitude is a clear sign that the result is sensitive to the noise model there. } 
\label{mcqcomp}
\end{figure}

It is important to understand this discrepancy.  We have already discarded CDPP as a useful measure of stellar variability in this context, and thus do not discuss it further. We therefore set out to reproduce the results of \citet{McQuillan2011}, whose methodology is relatively similar to ours. We were able to find new (similar) values for our two free parameters ($\delta I_0(0.5\rm{hr})$ and $N_0(0.5\rm{hr})$) that reproduce their results fairly closely. One difference (not important) is that their model implicitly takes $V^{min}$=0. Fig. \ref{mcqcomp} shows that our version of their model and our own model are very close fainter than about magnitude 13.5 (the divergence at brighter magnitudes is a direct result of the different values for $V^{min}$). One important difference lies in the way that we choose a value for the active Sun. Their approach is to take an average of the active Sun during two years approaching solar maximum in Cycle 23. Cycle 23 admittedly is not the cycle with the highest sunspot number recorded, but it is a typically active cycle. Using the average is one reasonable approach, but a caveat is that the Sun itself will yield an active fraction of 20\% (the fraction of solar points above this average). Our approach is to use the maximal solar variability to find the fraction of stars that are clearly more active than the Sun ever was in Cycle 23.  The caveat is that some stars that are in fact more active than the Sun could be in a quiet part of their cycle and lie below that limit. This, however, is largely dealt with in the histogram approach above. 

The active fraction is a sensitive function of the active solar value taken (a point which \citet{McQuillan2011} also make). We were not able to reproduce their value of 0.766 ppt for the average active Sun; we find a value just over 1.0 ppt for their stated time period of 2 years (approximately RJD 1400-2150), and their given value is close to the average for their entire dataset. Using their methodology with 1.0 ppt we do come very close to obtaining the same active fraction of 60\% that they get. The active fraction would be more like 90\% using 0.766 ppt, which shows its sensitivity to the solar level chosen in this range of levels. Using a maximal value of 2.0 ppt instead (which is exceeded by only 3 of the one-month range values in Fig. \ref{virgo}) reduces the active fraction to 30\%; taking a value of 2.5 ppt drops it to 20\% (the sensitivity of the active fraction decreases as the solar level rises above most of the noise). This is entirely consistent with what we found above using the histogram analysis on the bright stars. As discussed below, the activity level of 2.0 is high enough that the noise even for the faint stars does not dominate; we adopt this value for the rest of the paper. So far it appears that the primary source of disagreement between us and \citet{McQuillan2011} has very little to do with how we treat the noise (which is compatible) and a lot to do with how we choose the active level that characterizes the Sun. It is somewhat a matter of taste whether to use the average or extreme active Sun as a limit. We argue that it is more pointed to ask what fraction of stars are truly more active than the Sun ever gets, rather than comparing with the average active Sun, but one can quibble about that. The agreement of our choice with the histogram analysis (which partially deals with the cycle phase problem and does not depend as sensitively on the solar cutoff) leaves us convinced it is the better one.

\begin{deluxetable}{lcccccc}
\tablecaption{Fraction of Bright Stars More Active than Solar \tablenotemark{a}}  \tablenum{2}  \tablecolumns{8} \tablewidth{0pc}
\tabletypesize{\footnotesize}  
\label{fractab}
\tablehead{\colhead{ } & \colhead{late-F (1340) \tablenotemark{b}} & \colhead{G (974)} & \colhead{early-K (904)} & \colhead{mid-K (663)} & \colhead{late-K,M (1201)} & \colhead{late-K,M }\tablenotemark{c} }  
\startdata 
  & 6500-6000K & 6000-5500K & 5500-5000K & 5000-4500K & 4500-3500K &  \\
%\enddata  
%\startdata  
\hline
$R_{var}$(30 day) &  .17 &.28 & .41 & .64 & .91 & .63 \\
%$R_{var}$(30 day) &  .83 &.72 & .59 & .36 & .09 & .37 \\
%3 hr& .12 & .32	& .23	 & .38 &  .01 & .19 \\ % \tablenotemark{d}
6hr & .57 &.35 & .42 & .77 &  .96 & .61 \\
%6hr & .43 &.65 & .58 & .23 &  .04 & .39 \\
12hr & .49 &.32 & .38 & .60 &  .87 & .58 \\
%12hr & .51 &.68 & .62 & .40 &  .13 & .42 \\
24hr &  .41 &.32 & .38 & .58 & .86 & .59 \\
%24hr &  .59 &.68 & .62 & .42 & .14 & .41 \\
2day & .25 &.30 & .39 & .58 &  .87 & .61 \\
%2day & .75 &.70 & .61 & .42 &  .13 & .39 \\
4day & .16 &.30 & .42 & .63 &  .91 & .66 \\
%4day & .84 &.70 & .58 & .37 &  .09 & .34 \\
8day & .12 &.26 & .41 & .63 &  .90 & .60 \\
%8day & .88 &.74 & .59 & .37 &  .10 & .40 \\
\enddata
\tablenotetext{a}{The fraction of stars encompassed by solar activity is these numbers subtracted from unity. The value of $R_{var}$(30d) we use for the most active Sun is 2.0 ppt. }
\tablenotetext{b}{The number of stars in each sample is in parentheses.}
\tablenotetext{c}{Fraction of this group lying below three times the most active Sun in the noise model. }
 \end{deluxetable}

There is another source of disagreement, however, and that is the sample of  \textit{Kepler} stars used to make this comparison. \citet{McQuillan2011} use slightly different selection criteria to chose G dwarfs, perhaps the most important difference there is that their definition of G dwarfs (adopted from \citet{Ciardi2011}) extends down to 5300K (rather than our 5500K). That does (as we show below) tend to include more variable stars at the cooler end. In reproducing their results we restricted the sample to our temperature range. What is much more important, however, is the magnitude cutoff for the sample. We have argued above that when moving to stars fainter than $M_{Kep}$=14, the influence of noise becomes increasing problematic. We base our results for solar analogs on a sample that stays brighter than  $M_{Kep}$=12.5 (although in subsequent sections of this paper we relax the limit for cooler stars to retain a large enough sample). We can compute the active fraction for G dwarfs for various magnitude limits, and for active Sun values of $R_{var}$ of 1 and 2 ppt. For the stars with higher $R_{var}$ (which we adopted), the magnitude limit matters less. With magnitude limits of 13, 14, 15, and 16 the active fractions are 23\%, 21\%, 21\%, 26\% respectively. With the lower value of $R_{var}$ (which they adopted), however, the active fractions are 37\%, 38\%, 48\%, and 61\%. This is true even without including the stars from 5500-5300K (which would raise the fractions further). Thus we also see that \citet{McQuillan2011} would be in closer agreement with us if they simply restricted themselves to a sample that was brighter than $M_{Kep}$=14. They would then find that nearly two-thirds of the \textit{Kepler} stars are like the Sun. The result for the full sample and lower solar activity level is hardest to defend as intrinsically stellar, since it depends most on the apparent magnitude of the sample. 

There is one possible defense: if the fainter stars belong to a significantly different population than the brighter ones (since there is a difference in their distance from us). This would require the distant stars to be a substantially younger population (in the spirit of the arguments of \citet{Gilliland2011}). Their analysis used the results of Galactic population synthesis by TRILEGAL \citep{Girardi2000, Girardi2005} to argue that the Kepler field has a greater number of young stars than might be expected. However, the Kepler field lies at a Galactic latitude that is difficult for TRILEGAL to model accurately with respect to star counts. In the original iteration of TRILEGAL the counts at b=10 could be underestimated by a factor of 2 or more \citep{Girardi2005, Santerne2012}. Changes were made to TRILEGAL's population synthesis to bring the star counts at the latitude of the Kepler field into agreement with observation which we summarize briefly; see \citet{Girardi2005} for a detailed description. The slope of the luminosity function was changed by decreasing the contribution of the oldest disc components, which initially caused the local population to be too numerous and dominated by young stars. This was mitigated by implementing a hyperbolic secant squared vertical distribution for the thin disc. Finally, the star formation rate between 1 and 4 Gyr was (arbitrarily) increased by 50\%. While these changes were necessary to reproduce star counts at the latitude of the Kepler field, they all affect the age distribution of stars predicted by the model, rendering that substantially more uncertain. Unlike star counts, the age distribution at a given location in the Galaxy is generally not well known, and it is difficult to compare the ages in the synthesis model with observations. It would be rather coincidental if this special age distribution happened to produce the same rising variability at faint magnitudes as the influence of noise in the light curves does, so we do not find this a convincing alternative at the moment. 

\subsection{Variability on Different Timescales}
\label{sec:timevar}

Another way of comparing the Sun to the \textit{Kepler} stars is to make a differential comparison by utilizing the difference in variability between the quiet and active parts of the solar cycle and applying it to the \textit{Kepler} data. We can take the noise model in the previous section as a representation of the quietest stars at all magnitudes (we know the solar data can be even quieter in some cases). We can then implement a solar comparison by defining a $V_*^{act}$ for each t$_{bin}$ to be $V_*^{min}$ plus the difference in MDV($t_{bin}$) between the quietest and most active SOHO segments for a given timescale. The exact value of the quietest Sun does not matter, because it is a small number compared with the most active value (by factors of several at short timescales and more than an order of magnitude at timescales greater than 12 hours). Thus the desired difference is essentially just the most active value of MDV($t_{bin}$ for the Sun. We are thereby translating the difference from quiet to active variability on the Sun at each timescale to the stars at the same timescale, by preserving the lower envelope for the stars and adding the solar difference (or upper variability level) to that low stellar envelope. 

\begin{figure}
\begin{center}
\includegraphics[width=1.0\textwidth, height=6 in]{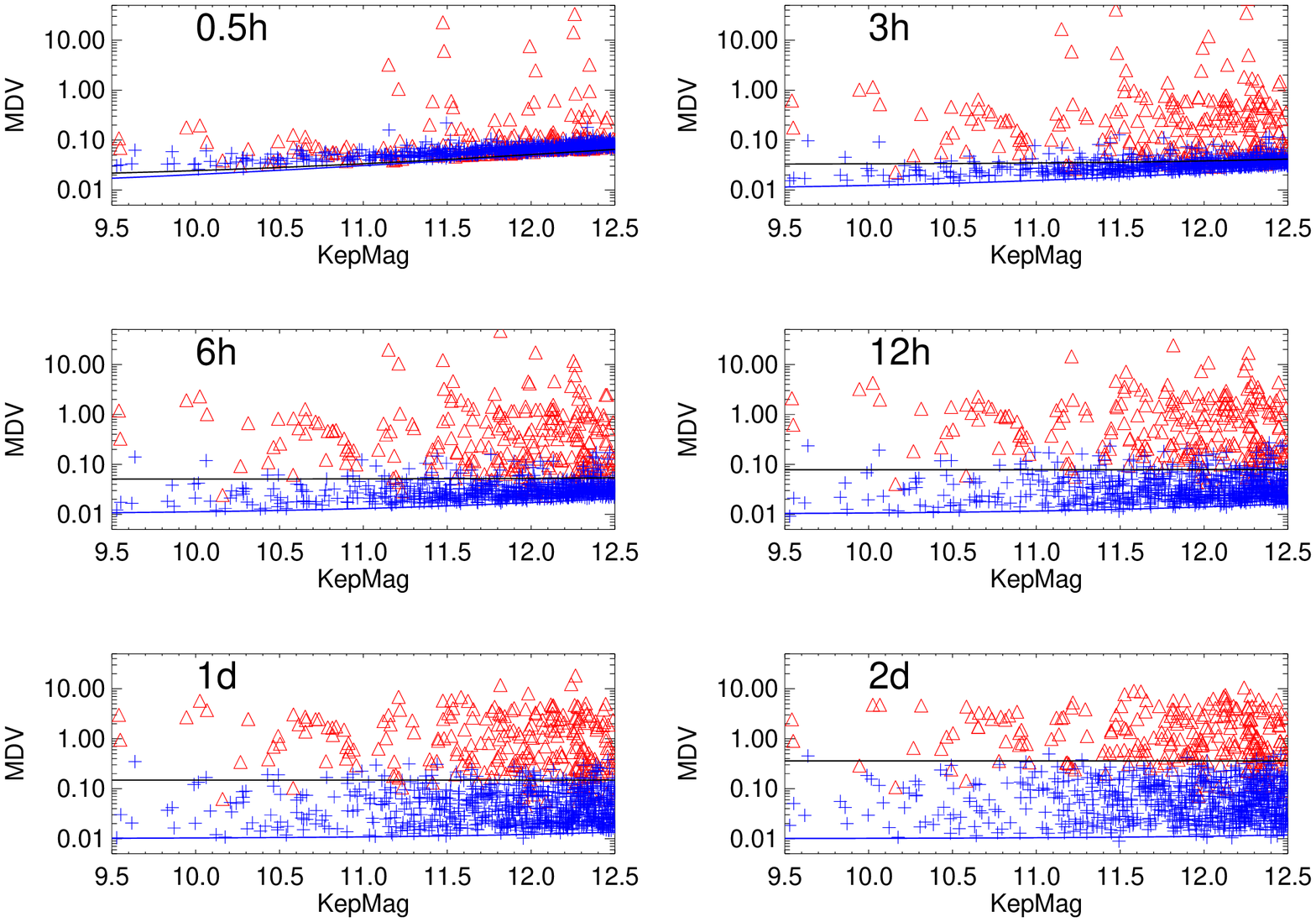} \\
%\subfloat[][]{\includegraphics[width=1.0\textwidth, height=2 in]{F6a.eps}} \\
%\subfloat[][]{\includegraphics[width=1.0\textwidth, height=2 in]{F6b.eps}} \\
%\subfloat[][]{\includegraphics[width=1.0\textwidth, height=2 in]{F6c.eps}} \\
\end{center}
\caption{The variability on different timescales of G dwarfs as a function of their brightness. The lower blue solid line is the minimum variability model based on the full \textit{Kepler} sample, while the upper black solid line is a predictive model based on SOHO data for the upper envelope expected in each plot for stars like the very active Sun. The blue plusses represent stars whose $R_{var}$ lie below that of the most active Sun and the red triangles are stars whose range exceeds the effective maximum of our SOHO data. The units of MDV in all figures are parts per thousand (ppt). }
\label{gdwf}
\end{figure}

We use this value of $V_*^{act}$ to represent what a star of a given magnitude would look like if it happened to have similar cycle amplitudes (range differences) as the Sun. This hypothetical star can be compared to what the stellar range distributions actually look like. Of course, the stellar sample covers a distribution of stellar parameters and ages rather than the unique solar values of these.  What we learn from this comparison is how well our hypothetical Sun-as-star predicts the MDV of stars of different brightness whose activity levels are within the solar range. The noise and activity models are shown with the lower solid (quiet) and upper solid (active) lines in Fig. \ref{gdwf}. The panels show MDV($t_{bin}$) as a function of $M_{Kep}$ for timescales up to 8 days. The \textit{Kepler} points are again distinguished by $R_{var}$: the blue plusses have $R_{var}$ up to our chosen maximum active SOHO value (2.0 ppt), the red triangles have $R_{var}$ greater than that. The fraction of all the points which lie within the solar range defined by the solid lines in Fig. \ref{gdwf} are given in Table \ref{fractab}. It is hard to estimate these fractions by eye from the figure, because a lot of the points overlap each other. The value of $V_*^{act}$ that we use for each case can be read in each panel of Fig. \ref{gdwf} and its analogs as the value where the upper horizontal line meets the y-axis.

The panel for 0.5\rm{hr} shows the same effect mentioned for Fig. \ref{rangemdv} -- that the quiet and active solar values are very similar to each other and lie below the \textit{Kepler} values for all magnitudes. There is also a lot of mixing between the various \textit{Kepler} $R_{var}$ (because MDV(0.5\rm{hr}) is not sensitive to magnetic activity). Much the same holds at 3hr, although the most active (high $R_{var}$) \textit{Kepler} points now rise above the others, and the model solar lines are separate at bright magnitudes (less so at fainter magnitudes because noise dominates there on this timescale). The active solar line caps the \textit{Kepler} points with solar-like $R_{var}$. The points are sorted by $R_{var}$ down to $M_{Kep}$=12.5 for timescales 6 hours or longer. The Sun is a good predictor of the relationship between $R_{var}$ and MDV($t_{bin}$) (or equivalently, between magnetic activity and differential photometric variability at a given timescale). For timescales of 6 hours up to 1 day, the red dashed line lies a little below the highest intermediate blue points, which means a few of the G stars with $R_{var}$ similar to the active Sun have variability on these timescales a little higher than the Sun.

We can re-assess yet again the ``active fraction'' of stars in the  \textit{Kepler} field. The previous subsection, and \citet{Basri2010}  and \citet{McQuillan2011} utilized $R_{var}$ as an activity diagnostic; here we can examine a full set of different timescales over 3 months for bright sample. It is simple to find the fraction of stars that are more variable than the active Sun at each timescale; we only have to count the points above the upper solid line in each of the panels of Fig. \ref{gdwf}. These fractions are given in Table \ref{fractab}. We discuss the detailed quantitative results in Section \ref{sec:othervar}, but they confirm the result of \citet{Basri2010} that at most a third of the comparable stars in the  \textit{Kepler} field are more active photometrically than the active Sun for almost of the timescales considered here. This fraction is what would be expected by simply assuming a uniform star formation rate and well-mixed sample and noting that the Sun is about one third the age of the Galaxy \cite{Batalha2002}. We do not claim this ``back of the envelope'' estimate is actually the correct explanation (it is definitely oversimplified), but it does empirically yield roughly the right answer. 

We have found active fractions employing several methodologies that are between 20-33\% (depending on the timescales and assumptions used). They imply that the Sun has a ``normal'' level of magnetic activity compared to the \textit{Kepler} sample. We find no empirical evidence in the \textit{Kepler} data for an excess of young active solar-type stars near us, nor is the Sun unusually photometrically quiet compared to its neighbors. That is perhaps not surprising, given similar results for CaII (\cite{Henry1996}). We have explained in detail above why these results are at variance with those of \citet{McQuillan2011}. These conclusions are also in some conflict with \citet{Gilliland2011}, who found (as we do) that the \textit{Kepler} stars in the $M_{Kep}$=11.5-12.5 range are more variable than the Sun at timescales of 6 hours, using an analysis based on CDPP. We have argued above that MDV is the more appropriate measure for assessing the stellar contribution (and it is certainly more straightforward to understand). We conclude here that the bright \textit{Kepler} solar-type stars are more variable than would be predicted by the average Sun from SOHO data because the noise not intrinsic to the stars in \textit{Kepler} data is still the dominant factor at a timescale of 6 hours for $M_{Kep}$=11.5-12.5. The behavior of the bright sample at timescales 12 hours or greater indicates that most of them are quite comparable to the Sun in photometric variability. We know of no physical mechanism that would make solar-type stars more photometrically variable than the Sun at timescales of a few hours but not at timescales of half a day or longer. It is possible that the PDC-MAP reduction process still does not fully preserve stellar variability, but it is unlikely that such problems would yield consistent results over the broad set of timescales we have considered here. We do see some indications of such a problem at the longest timescales (the lower scatter of  \textit{Kepler} at 8 days and the fact the Sun has quieter $R_{var}$ at the extreme). Our results call into question the age distributions in the galactic models used by \citet{Gilliland2011}. It will require substantial additional work to confirm or reject this possibility.
 
 \section{Variability of Other Spectral Types}
 \label{sec:othervar}
 
 We now carry out the same analysis as in the last section for samples of main sequence stars in other temperature bins. In each case we take a bin with a temperature width of 500K and choose faint magnitude limits to obtain a sample of the brightest stars of approxmimately the same size (a thousand or so) in that temperature bin. We have chosen simple magnithde limits rather than a strictly controlled sample size somewhat arbitrarily (because we are more interested in the noise-magnitude dependence and we are not comparing strict sample statistics). We do not repeat the comparison by $R_{var}$, but concentrate on MDV($t_{bin}$) versus $M_{Kep}$ for each temperature bin. As before, blue plusses represent stars with values of $R_{var}$ within the range displayed by the Sun, while red triangles have $R_{var}$ greater than any solar points.

\begin{figure}
\begin{center}
\includegraphics[width=1.0\textwidth, height=6 in]{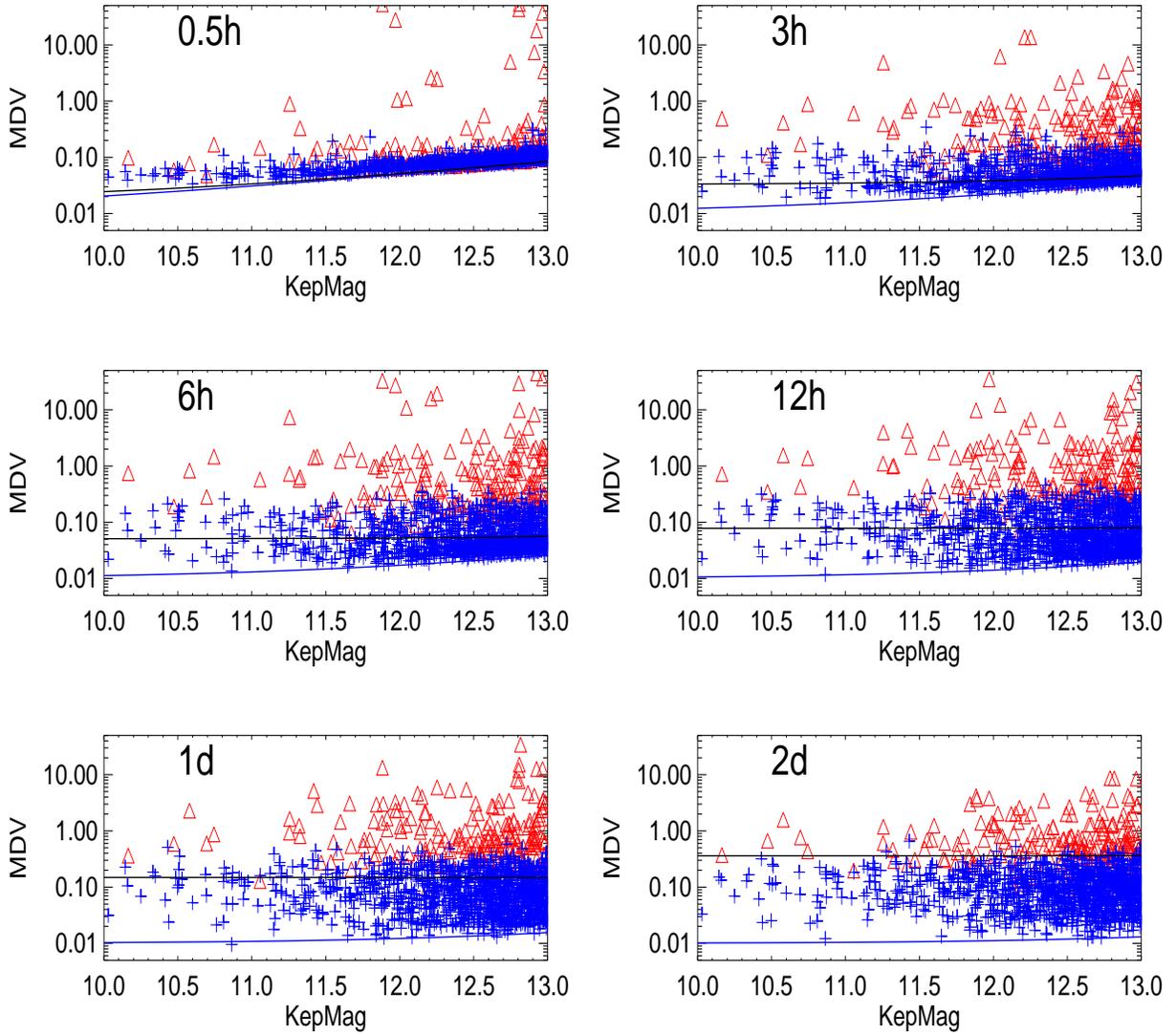} \\
\end{center}
\caption{The variability on different timescales of late-F dwarfs (6500-6000K) as a function of their brightness. The meanings of the symbols and lines are the same as for Fig. \ref{gdwf}. }
\label{fdwf}
\end{figure}

We start with the bin just hotter than our solar-type sample, namely 6500-6000K. We extend the magnitude limit down to $M_{Kep}$=13 which yields a sample of 1340 stars. Fig. \ref{fdwf} shows the results, which are not that different from Fig. \ref{gdwf} for the G stars. A larger fraction of the stars (34\%) are in the intermediate range which matches the active Sun than for G stars (19\%). These are taken from a reduced fraction of stars with $R_{var}$ above that of the active Sun (the definitions of quiet, intermediate, and active are as in the last section). Table \ref{fractab} shows that only 17\% of the late-F stars (these spectral type names are only approximate; the bins are defined by KIC effective temperature) have $R_{var}$ greater than the most active Sun (compared with 28\% for the G stars). This means that these late-F stars  do not have as many examples of strong photometric variability as the G stars; more of the late-F stars look like the active Sun in $R_{var}$ while more of the G stars are even more active than that. At timescales from 6-24 hours there are more late-F stars that have $R_{var}$ similar to the active Sun but are more variable at these timescales than the active Sun. The fraction of F stars that are more active than the Sun (as defined by the upper solid line) is higher for the late-F stars than the G stars up to 2 days, but begins a dramatic drop at longer timescales (Table \ref{fractab}), down to only 12\% at 8 days. This can be seen in Fig. \ref{fdwf} where the top of the cloud of blue plusses moves from above to below the upper solid line as the timescales increase. Thus we can say that F stars are quieter than G stars at longer timescales, but more active on the timescale of about a day.
 
\begin{figure}
\begin{center}
\includegraphics[width=1.0\textwidth, height=6 in]{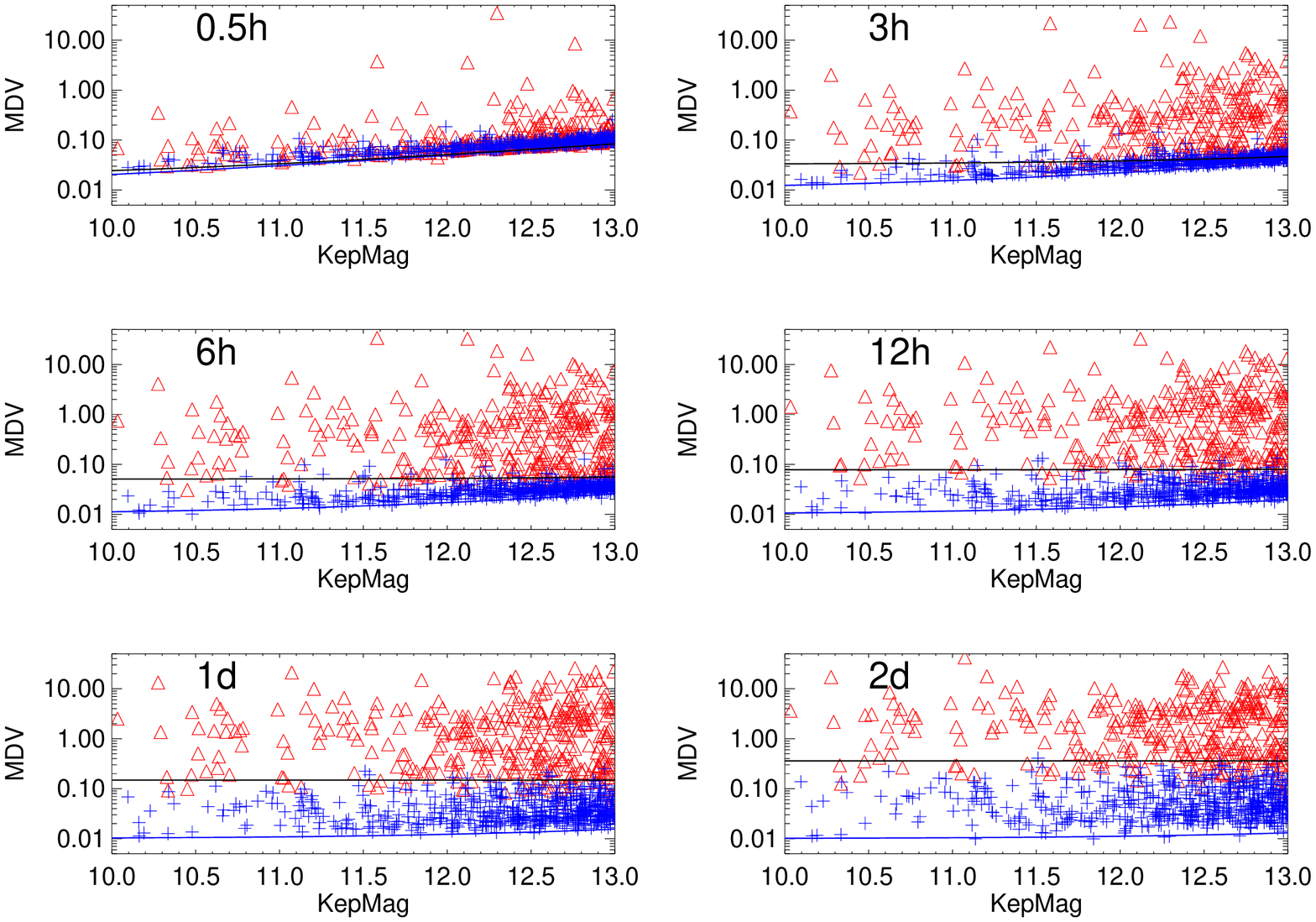} \\
\end{center}
\caption{The variability on different timescales of early-K dwarfs (5500-5000K) as a function of their brightness. The meanings of the symbols and lines are the same as for Fig. \ref{gdwf}. }
\label{k0dwf}
\end{figure}

Considering the temperature range just below the G stars, namely 5500-5000K, we obtain a sample of 904 stars from a magnitude limit of $M_{Kep}$=13. These early-K stars are shown in Fig. \ref{k0dwf}. We see a somewhat different distribution of $R_{var}$ in this sample; the fraction of stars more photometrically variable than the Sun is higher and the fraction of quieter stars is lower (Table \ref{fractab}). The active fraction of early-K stars in $R_{var}$ than the Sun is 41\% while it was 28\% for the G stars. The fractions of very active stars at different timescales are also higher for the early-K stars, by amounts ranging from 6\% percent more active at half a day to 15\% more active at 8 days. The three samples we have discussed so far are most similar at 1 day timescales, with the G stars being the least variable of them at that timescale. The active line tends to be at the top end of blue points for the early-K sample (Fig.\ref{k0dwf}), suggesting that for these stars the Sun is a good predictor of variability at different timescales if one compares their $R_{var}$ to solar $R_{var}$.

\begin{figure}
\begin{center}
\includegraphics[width=1.0\textwidth, height=6 in]{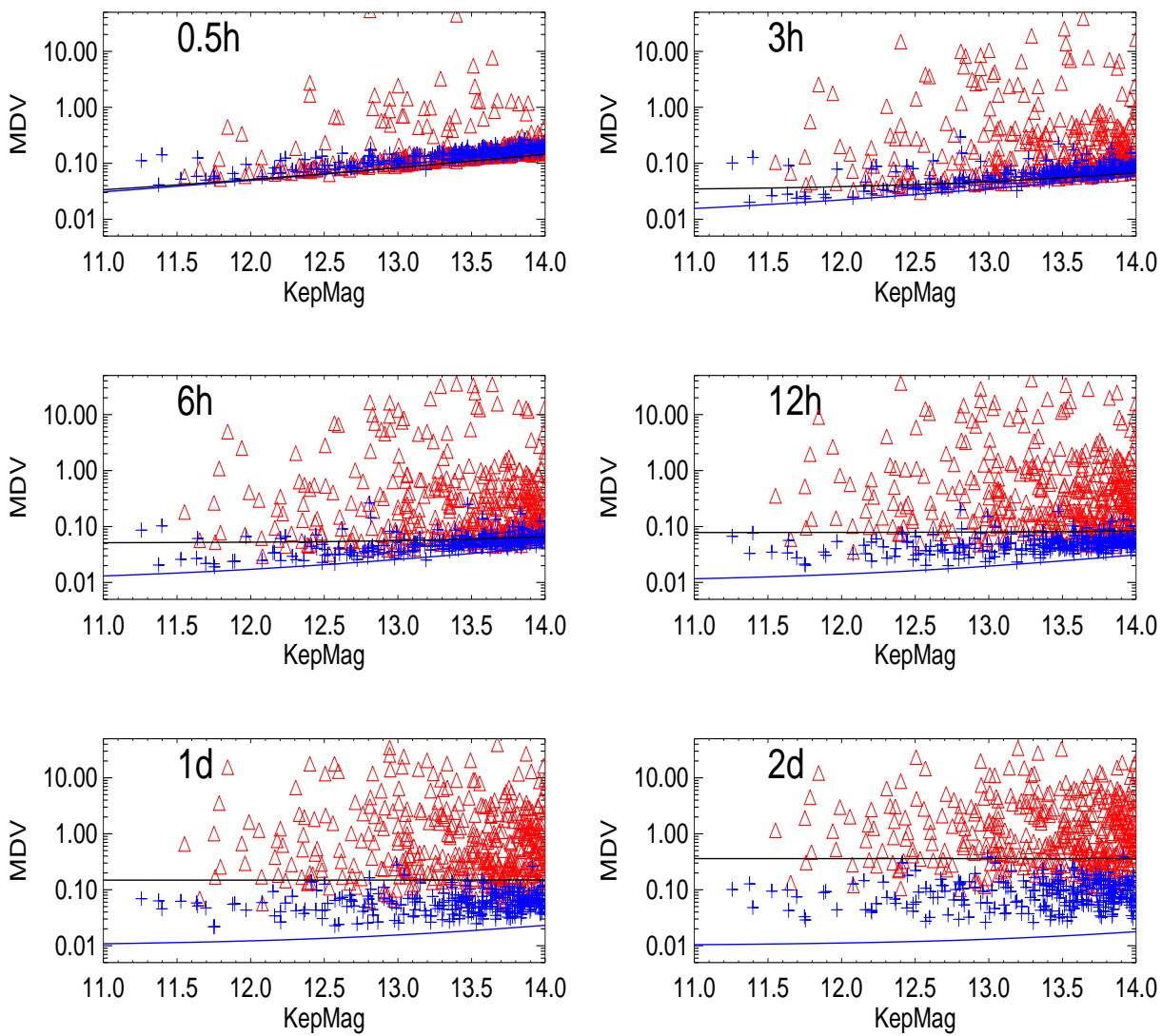} \\
\end{center}
\caption{The variability on different timescales of mid-K dwarfs (5000-4500K) as a function of their brightness. The meanings of the symbols and lines are the same as for Fig. \ref{gdwf}. }
\label{k5dwf}
\end{figure}

Moving to the next cooler bin (5000-4500K) we have to reach down to $M_{Kep}$=14 to obtain a sample of 663 stars of mid-K stars. In this case we prefer to analyse a smaller sample rather than cross the limit where we have shown that noise is clearly becoming influential. This set is shown in Fig. \ref{k5dwf}. At the shortest timescales these points are closer to the lower solid line as expected, since instrumental noise is playing a a larger role for these fainter stars. The fraction of stars more active than the active Sun (in $R_{var}$) is also larger at 64\% (compared with 41\% for early-K stars). As with the early-K stars, the stars which are within active Sun $R_{var}$ also lie mostly within the predictive line set by the behavior of the Sun at each timescale. There is a reversal of the situation for late-F stars at 2 days, in that some of the active (red) points lie below the solar active solid line (meaning that these stars which are more variable on a one-month timescale are less variable on a 2 day timescale). Most of the more active stars, however, exhibit greater differential variability on all timescales than the Sun does. The fact that the points lift off the lower line at longer timescales also implies that stellar variability is important even at the faint end (ie. these stars are generally more variable than the quietest stars in the full sample). The cooler stars have short timescale photometric behavior that is similar to the Sun, when one uses $R_{var}$ (determined over a month) to predict the shorter term behavior. Thus, general activity studies based just on $R_{var}$ will typically yield the same qualitative results as more detailed studies based on shorter timescales.

\begin{figure}
\begin{center}
\includegraphics[width=1.0\textwidth, height=6 in]{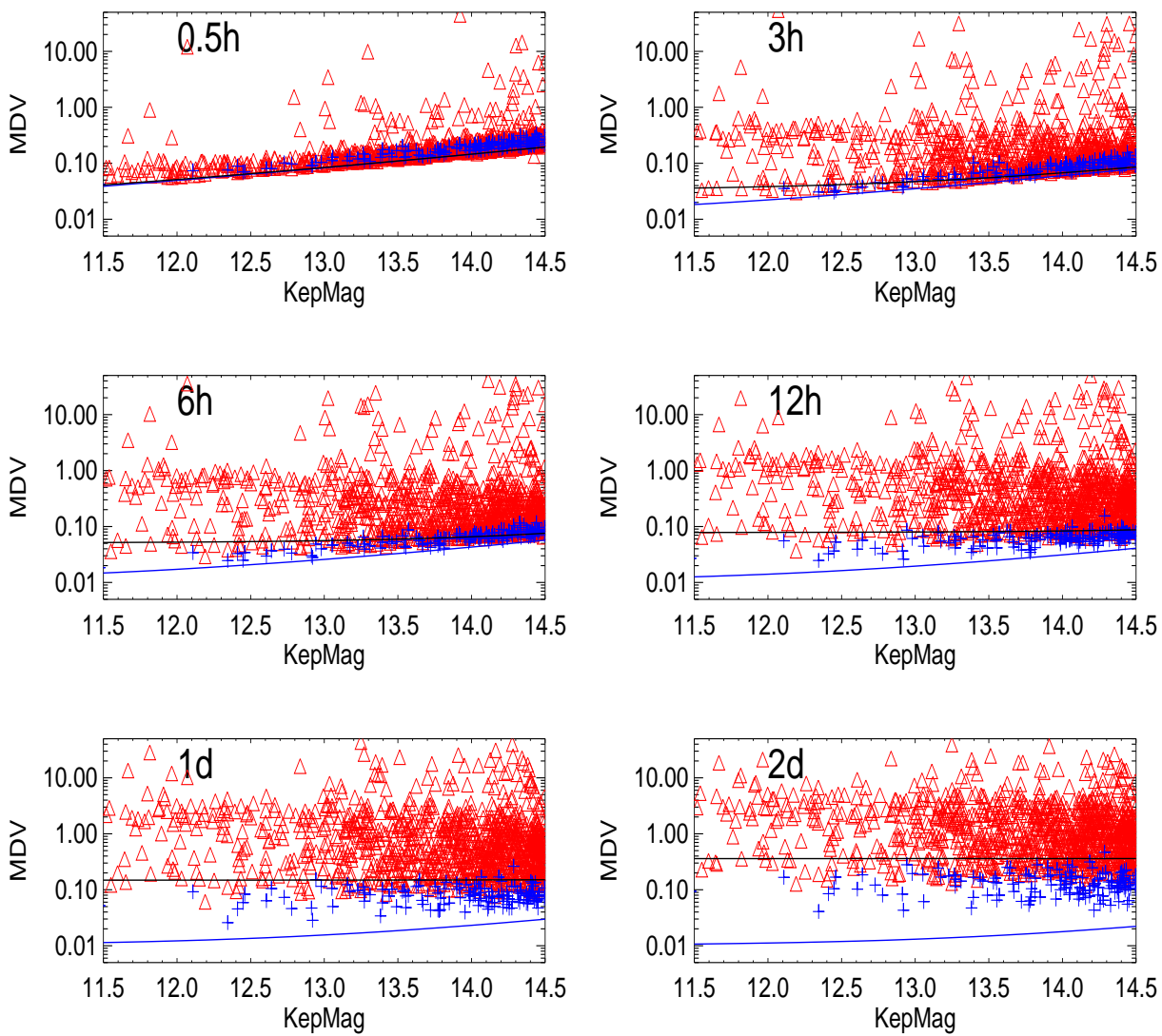} \\
\end{center}
\caption{The variability on different timescales of late-K (4500-4000K) and early-M dwarfs ($<$4000K) as a function of their brightness. Approximately 30\% of these stars are early-M dwarfs. The meanings of the symbols and lines are the same as for Fig. \ref{gdwf}. }
\label{mdwf}
\end{figure}

Our coolest bin covers temperatures less than 4500K (still for gravities of log(g)$\geq 4.3$). We must extend further down to $M_{Kep}$=14.5 to produce a sample of 1252 late-K,M stars (30\% of them are early-M dwarfs). The other feature of note for the late-K,M stars is a clear band of excess population at the 3 hour and longer timescales, at MDV($t_{bin}$) values of 0.4, 0.8, 1.6, 2.5, and 4 ppt respectively. Alternatively, one could say that there is a concentration of stars at these values of MDV, a lesser population at the bottom envelope, and a deficit of stars in between. This holds over a span of magnitudes (down to about $M_{Kep}$=13), and is only present in this coolest temperature bin. We have filtered out giants using KIC gravity for our bright dwarf sample in Fig. \ref{mdwf}, but it is rather suspicious that we see these bands clearly only in the effective temperature bin where giants would be most common. Could they represent a contamination of lower gravity stars, or equivalently, errors in KIC gravities? \cite{Basri2011} showed that it is possible to identify giants by their photometric signature; they tend to have lots of periodogram peaks with relatively similar power (they are quasi-periodic) and timescales of variability that lead to many zero-crossings per week in a \textit{Kepler} light curve. We performed such a filtering on our sample, demanding that there be at least 10 periodogram peaks that are greater than a tenth the power of the highest one, that there be more than 30 zero-crossings in the light curve, and that R$_{var}$ be less than 10.  This produced only 51 obvious cases of giants based on the light curve properties, which we then excluded, reducing the count to 1201 stars. 

\begin{figure}
\begin{center}
\includegraphics[width=1.0\textwidth, height=6 in]{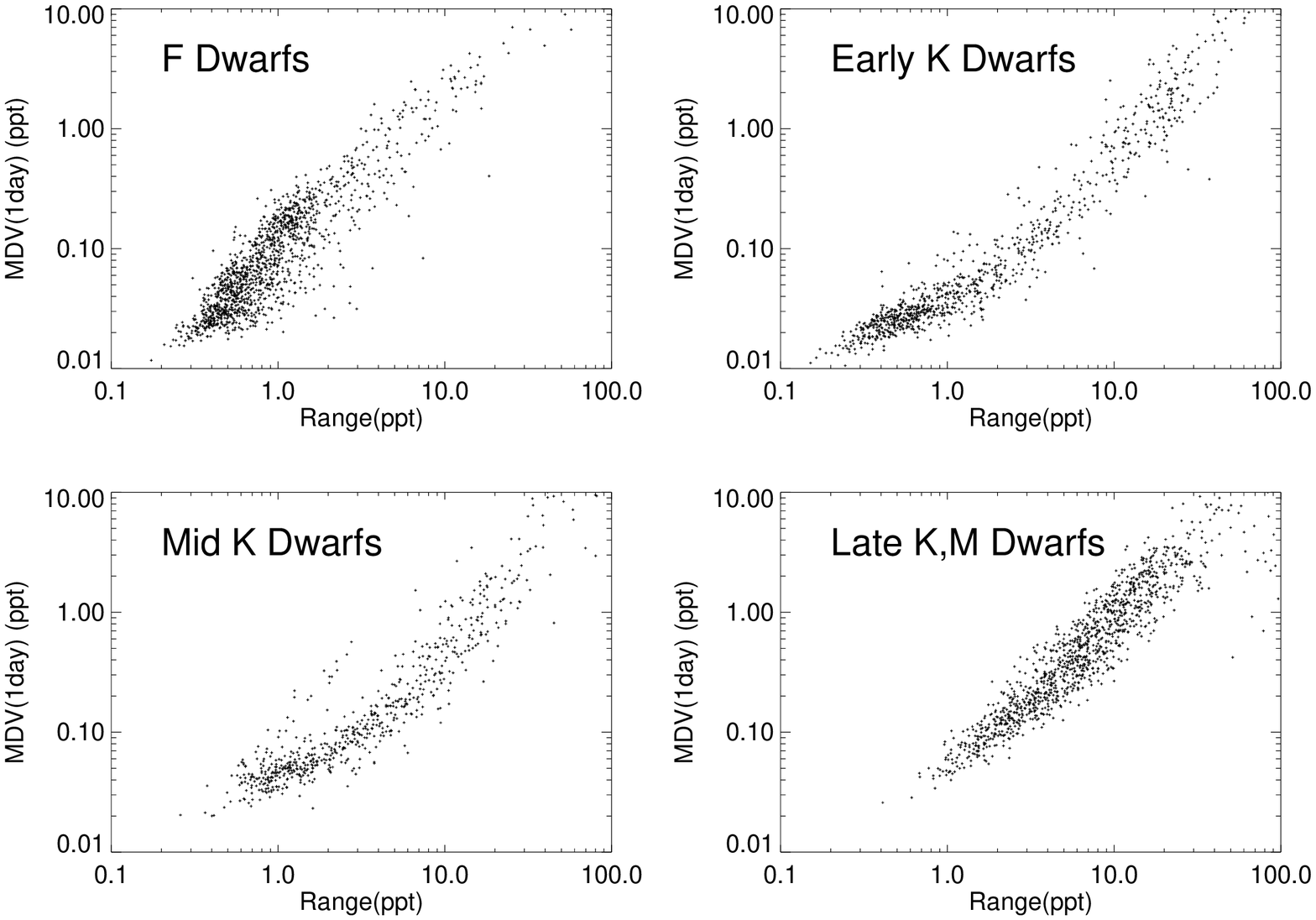} \\
\end{center}
\caption{The correlation between the  MDV(1-day) values for different spectral types, and their $R_{var}$ values for one month. These correlations are fairly tight, and hold for all the samples (with subtle shape changes and different spans).}
\label{rmcorrel}
\end{figure}

It is known that the KIC gravities have errors which report giants as dwarfs. \cite{Mann2012} discuss this point in detail, and have used medium resolution spectroscopy of 382  \textit{Kepler} targets to explicitly test the veracity of KIC gravities. They find that 96\% of the stars in their sample brighter than $M_{Kep}$=14 are giants; a much greater fraction than in the KIC. We have looked into the relevance of that result for our sample of cool stars. It turns out that we have in place several filters that already removed most giants of the sort they identify. \cite{Mann2012} selected their sample using photometric filters rather than KIC stellar parameters. It turns out that more than two-thirds of the stars that they spectroscopically classify as giants are actually unclassified in the KIC. Because we only used stars with KIC gravities of log(g)$\geq$4.3  we automatically exclude all unclassified stars. Of the 92 KIC classified stars in the \cite{Mann2012} sample which they identify spectroscopically as giants, 68 of them are also classified as giants in the KIC. We have 21 of the remaining 24 stars in our Q9 sample, which the KIC lists as dwarfs but \cite{Mann2012} find are giants. Thus, despite the dramatic conclusion that almost all of the Mann stars brighter than 14 are giants, we do not expect that more than a small percentage of our sample are likely to be giants given how we have chosen them. The remaining 21 stars are hardly a representative sample either; all but two of them are brighter than $M_{Kep}$=12.5 and most are brighter than 11. None of these have as rapid variations as typical red giants (see Fig. 5 in \cite{Basri2011} for examples). The light curves for 16 of them are shown in Figure \ref{lcsamp}. The zero-crossings are a day or two apart or longer, and a few of them are very slowly varying (and have higher amplitudes). It is possible these are even higher luminosity stars (taking the \cite{Mann2012} assertion that they are not dwarfs as true). It is a bit chastening to realize there might be this much diversity in giant light curves, although these are not typical red giants. 

\begin{figure}
\begin{center}
\includegraphics[width=1.0\textwidth, height=6 in]{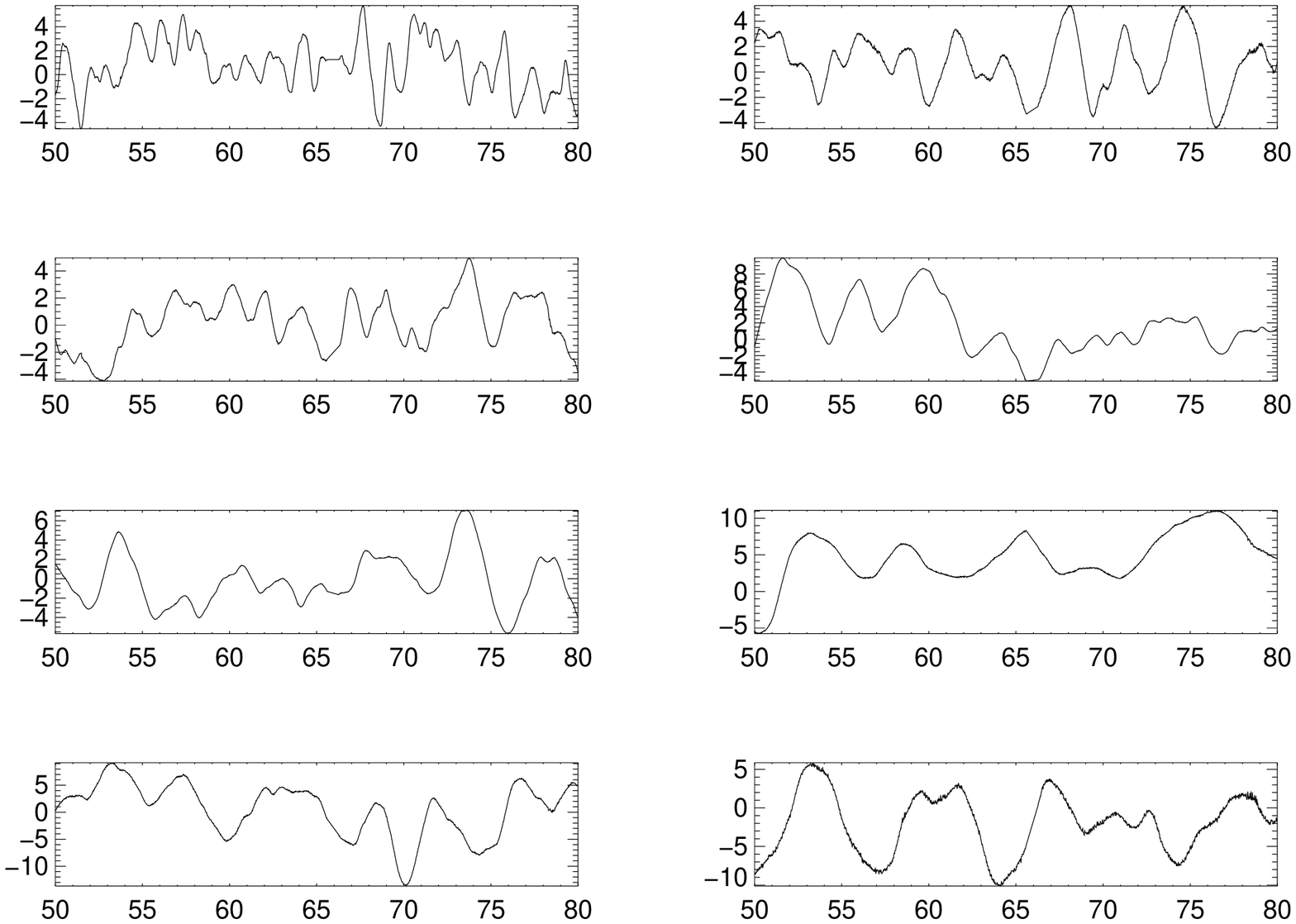} \\
\end{center}
\label{lcsampa}
\end{figure}

\begin{figure}
\begin{center}
\includegraphics[width=1.0\textwidth, height=6 in]{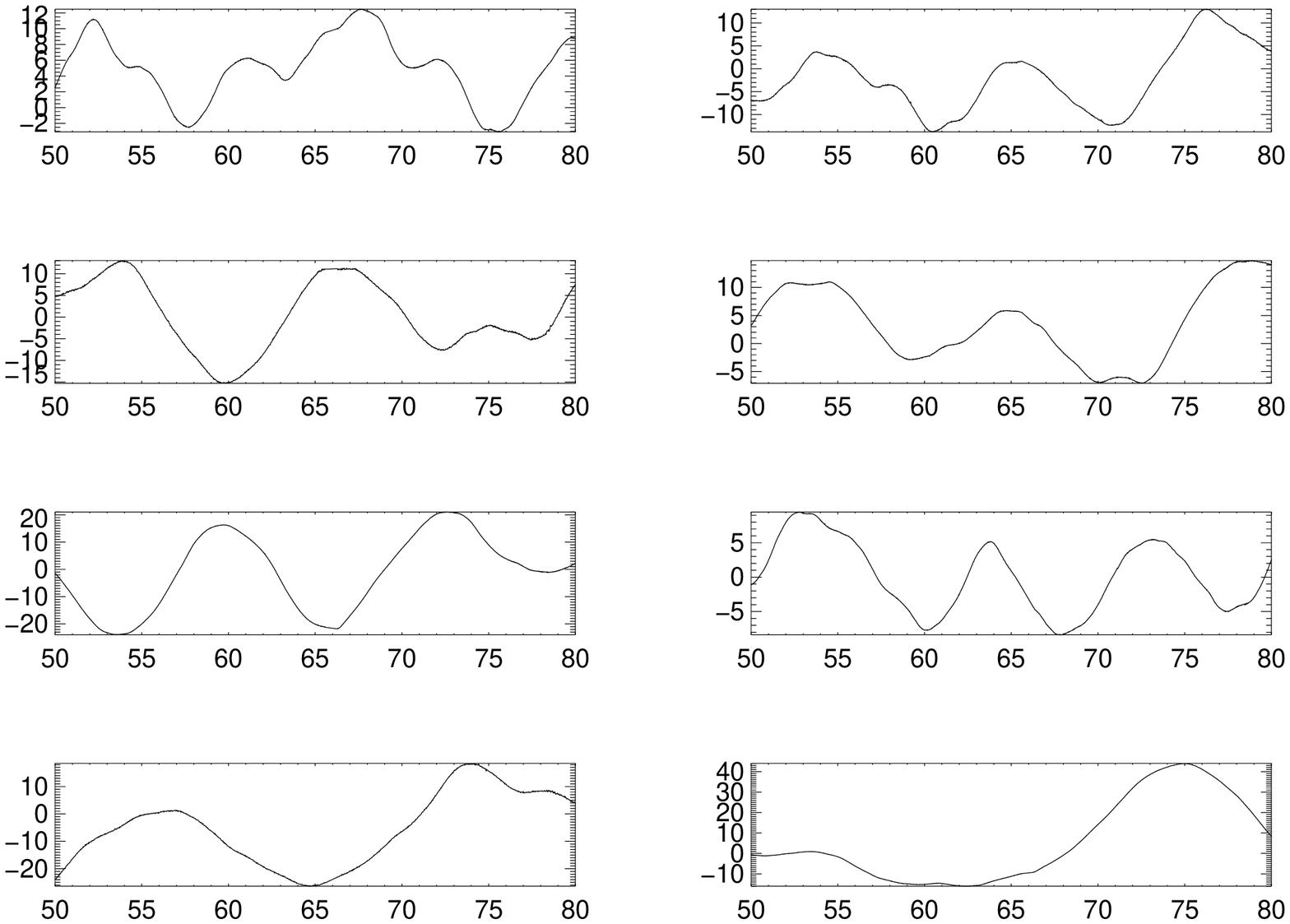} \\
\end{center}
\caption{The light curves over a month of stars which \cite{Mann2012} find are giants but which have KIC gravities suggesting they are dwarfs (and so which are included in our sample of cool dwarfs). Note that the first set of stars have variability which is somewhat slower than typical red giants and not particularly periodic, and the second set of stars is largely cases that are both higher amplitude and much slower than expected for red giants, as well as more periodic. Almost all these stars are at the very bright end of the sample; perhaps they are very luminous. The KepIDs for the upper panel are 7095218, 4385594, 1160867, 1026895, 8417203, 2424191, 3110253, 7505113 and for the lower panel they are 12155015, 9143855, 3964647, 5784204, 3218009, 11099165, 3964632, 3222519.}
\label{lcsamp}
\end{figure}

\begin{figure}
\begin{center}
\includegraphics[width=1.0\textwidth, height=6 in]{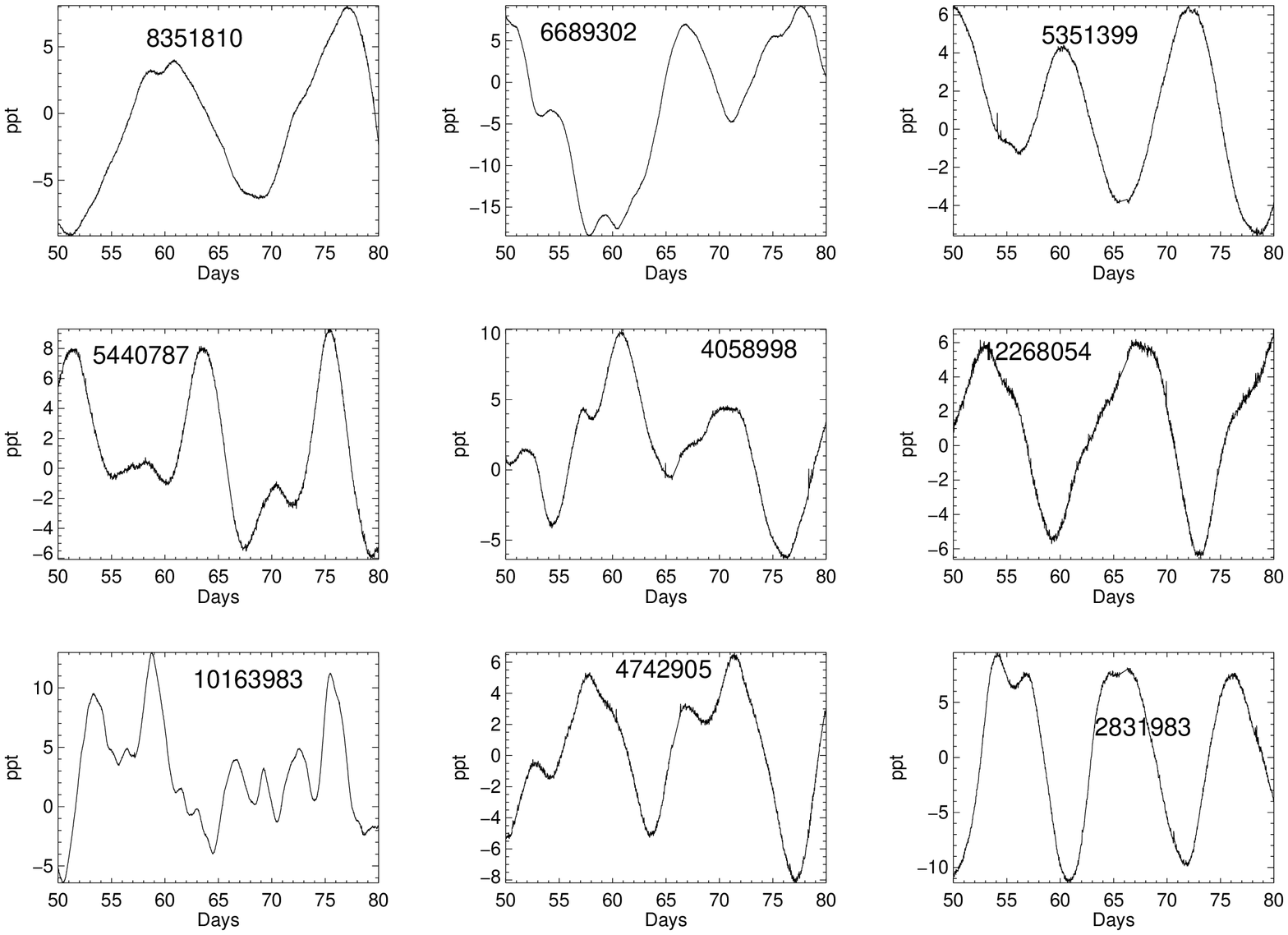} \\
\end{center}
\caption{The light curves of stars which lie within the band of excess population in our sample of cool dwarfs that can be seen in Fig. \ref{mdwf}. Many of these cases show clear periodicity on timescales of a week or more, including secondary slightly evolving features as are typically seen in solar-type spotted stars. The two which look more like the cases in Fig. \ref{lcsamp} are 6689302 and 10163983; they are less obviously periodic. }
\label{lcset}
\end{figure}

\begin{figure}
\begin{center}
\includegraphics[width=1.0\textwidth, height=6 in]{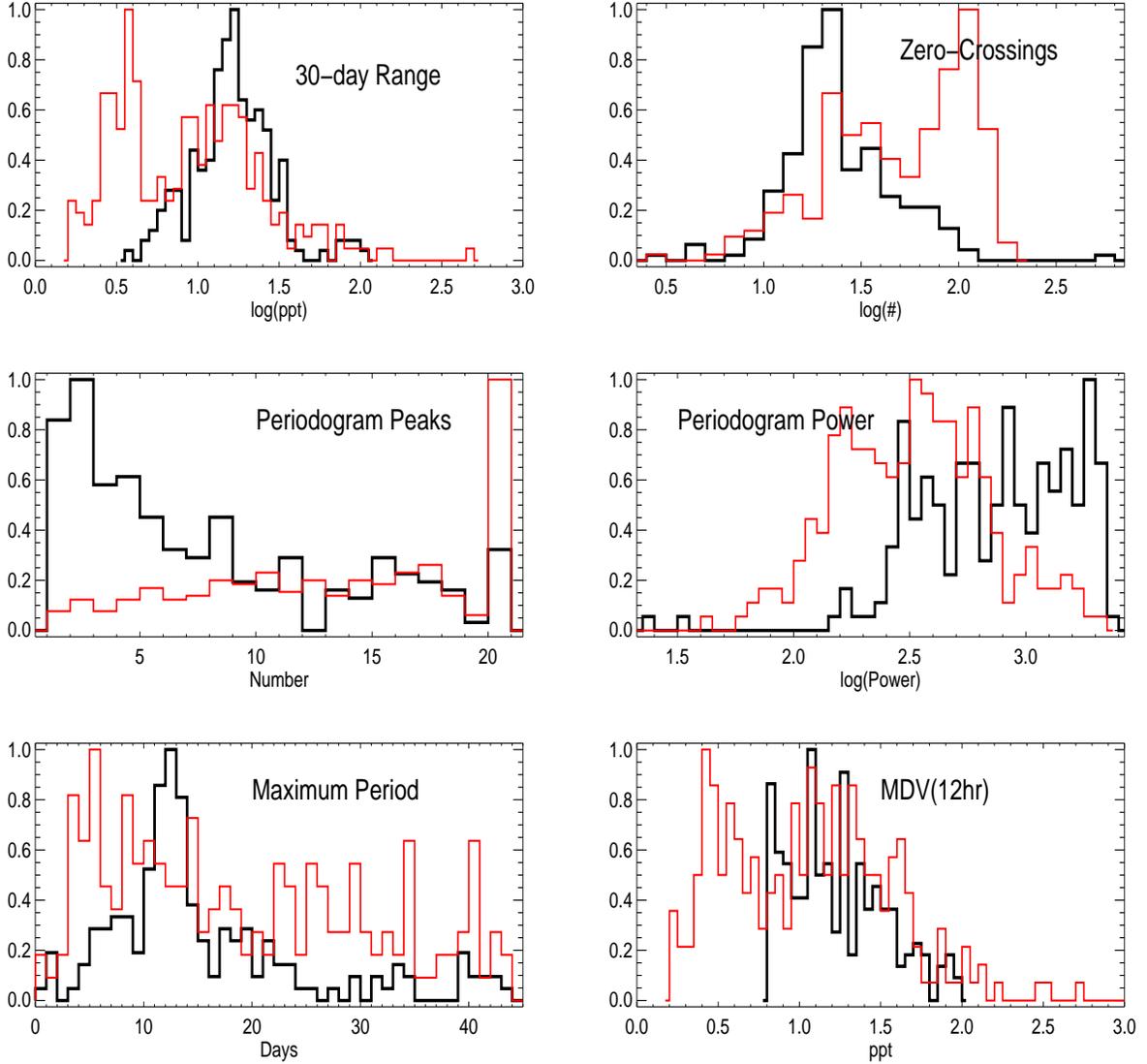} \\
\end{center}
\caption{A set of representative diagnostics of the light curves of the ``band'' stars in our sample (see text) shown as the thicker histogram, and KIC giant stars shown with the thinner red histogram. The first two diagnostics have to do with the amplitude and frequency of variability, and the second two have to do with the extent to which the variability is periodic. In all cases the ``band'' stars behave differently than the giants, although there is some overlap in their properties (and the appearance of the light curves). }
\label{gdcomp}
\end{figure}

We next examine the 129 light curves in the MDV(12 hr) case where the value of MDV lies between 1.0 and 2.0 (the location of the excess band). They look, for the most part, like spotted stars with appropriate amplitudes, clear periods of reasonable lengths, and evolving spot features. A small minority of them do look like the cases in Fig. \ref{lcsamp}. It is interesting that  seven-eighths of the stars in the band brighter than $M_{Kep}$=13 have KIC temperatures under 4000K (while this fraction is two-thirds for the whole band). The band is more obvious for brighter stars, which might be expected if they were giants, but the KIC temperatures are low for red giants. A representative sample is shown in Figure \ref{lcset}. In order to better understand to what extent we are looking at giants misclassified as dwarfs, we also extracted a sample of 261 bright giants, classified as such by the KIC (KIC gravities less than log(g)=3.8). Figure \ref{gdcomp} shows a comparison of the ``band'' stars with those giants on several dimensions of light curve properties; these diagnostics were discussed by \citet{Basri2011}. The band stars have typically higher ranges than the KIC giants by a factor of a few, although there is a secondary peak of giant ranges that are similar to the band stars. The band stars also show fewer numbers of zero-crossings by a factor of several (again with a secondary giant peak in the same place). 

The band stars are substantially more periodic than the KIC giant sample. When comparing the number of Lomb-Scargle periodogram peaks within a tenth of the highest one, the band stars concentrate at a low fraction of such cases, while the giants concentrate at the (self-imposed) limit of at most 20 such peaks that we tabulate. It is not surprising, therefore, that the highest power peaks are substantially higher in general for the band stars than the KIC giants. We also examined the histograms of the strongest periods for these two samples, now restricted to values of $R_{var}$ between 5 and 50 ppt (where most of the band stars lie). The band stars show a distribution peaked at 13 days with a sharp fall-off below 10 days and a steady decrease to longer periods. The giants, on the other hand, have a rising distribution, with an excess below 10 days and above 25 days compared to the band stars. Perhaps most striking is the difference between the MDV(12hr) distributions for the two samples; the band stars are restricted to a narrow range (by definition) in MDV(12hr) but the giants are substantially more dispersed. All of these indicators provide clear evidence that the band stars exhibit a significantly different (though overlapping) set of relevant light curve characteristics. 
 
We also looked at the set of stars from our sample with lower MDV(6 hr). These also look like spotted stars, but with smaller amplitudes and longer periods. By contrast, the 68 stars with MDV(6 hr) greater than 1.0 have both high amplitudes and very short periods (some less than a day). A few of them look like hot pulsators, which may indicate target contamination or another puzzle (no cool star should behave in this way). The totality of these somewhat puzzling results hints that there might be a bimodal distribution in the population of M dwarfs in the \textit{Kepler} field. They appear to be dominated by young active stars with something of a dearth of intermediate age stars (at least at these brighter magnitudes) and a less surprising small population of old stars. \cite{West2008} have shown that nearby samples of M dwarfs suffer from such a bias; they find that one needs to look out to magnitudes of fainter than 16th for early M dwarfs (or at least 100 pc above the galactic plane) before picking up a significant enough old disk population to counterbalance the nearby young disk population. The  \textit{Kepler} sample does not extend out that far. We are not confident in the suggestion of a bimodal distribution at this point, however, and feel that a better grasp of the extent of contamination by stars off the main sequence would be required to gain that confidence.

The results from our ``bright'' sample are shown in Fig. \ref{mdwf}. It is fair to say that the faint end of the sample is in danger of being affected by noise based on what we found in Section \ref{sec:compsun}). Examining Fig. \ref{mdwf} by magnitude shows little change in the qualitative behavior of the bottom envelope at the various timescales, however. What is happening is that while the faint envelope of \textit{Kepler} stars are indeed contaminated by noise, this particular sample of stars has excess stellar variability compared with the typical temperature of the \textit{Kepler} stars, so the effects of noise are not yet important in this coolest temperature bin. The discussion for these stars is much like that for the mid-K stars, except more so. The fraction of them which have $R_{var}$ greater than the active Sun is now 91\%, and almost none of them are less variable than the active Sun. Their variability lifts them off the lower line beginning at 12 hour timescales even for magnitudes that should be significantly affected by noise. The bottom of the cool star envelope stops being dependent on magnitude. That indicates that the stars are truly more photometrically variable (it is also obvious from examination of individual light curves, which are not dominated by noise). 

Since one might expect the spot contrast to actually be less for these very cool stars than for solar-type stars, this excess variability must be due to some combination of a great deal more spotting and perhaps a greater influence of bright (plage) regions on these stars. The latter would naively be expected from a larger contrast between plage temperatures and the cooler photospheres, although that may well be mitigated by the smaller sizes of the granules (the plage effect depends on the unmasking of the sides of granules by intergranular magnetic fields). It is still true that the few stars with $R_{var}$ no greater than the active Sun have variability at all timescales that is also no larger. One striking feature throughout the spectral classes is that the MDV($t_{bin}$) values for all the timescales not affected by noise are reasonably well-correlated with $R_{var}$, as can be seen in Figure \ref{rmcorrel}. This is despite the spread in rotation rates that must be present, and the changing levels of spot and plage coverage, as well as their contrasts with the photosphere. The correlations do show subtle differences that may be caused by these other factors, but generally one can predict the various variability values with one measurement (eg. $R_{var}$(1 month)).

\section{Conclusions}
\label{sec:conclude}

We have studied the variability of main sequence stars with convective envelopes at different timescales, ranging from half an hour to 3 months. This is done utilizing \textit{Kepler} data reduced with the new PDC-MAP pipeline, and some simple methods of characterizing variability. The first of these is $R_{var}$, which the difference between the 5\% and 95\% brightest differential intensities in a light curve over a specified timespan (either a month or a quarter in this paper); this is becoming a standard way of characterizing \textit{Kepler} stellar variability. Here we introduce a second metric, the median differential variability MDV($t_{bin}$). It is defined as the median absolute difference between all successive bins of a light curve that has been binned by a given timescale. Thus it is very akin to a median derivative of the differential intensity on a given timescale over a light curve. This variable can be used to characterize either noise (especially on very short timescales) or stellar variability at a timescale of interest. We consider time bins from half an hour to 8 days. We look at these two diagnostics for samples of the brightest \textit{Kepler} targets in a set of temperature bins with width of 500K ranging from 6500K down to 3500K (the coolest bin is 1000K wide). The magnitude limits are set so that we have of order a thousand stars in each sample. 

One goal of this paper is to compare the solar-type stars in the bright \textit{Kepler} sample to our Sun. There have been conflicting claims as to whether the Sun is typical or quiet compared to solar-type stars within a kiloparsec. The \textit{Kepler} light curves for solar-type stars are somewhat more variable than predicted before launch, in the magnitude range of greatest interest ($M_{Kep}$=11-12.5) and on a timescale of 6 hours. There has been some question whether that is due to the instrument (photon or other noise sources) or intrinsic stellar variability. In order to make the comparison with the Sun, we develop a simple 3 parameter noise model for \textit{Kepler} that is well matched to the lower envelope of variability for all stars and timescales. This model enables us to translate results from SOHO to a \textit{Kepler} scale.

We show that the SOHO data on the Sun is substantially less noisy than \textit{Kepler} at the long cadence timescale (half an hour). The characteristics of the \textit{Kepler} variability show the influence of noise at short timescales, and we  illustrate at what magnitudes and timescales stellar variability begins to dominate over noise. Timescales shorter than 12 hours seem to be clearly affected by noise, while longer than that the \textit{Kepler} and solar variability show similar spreads, and clear correlations with variability on one-month timescales. The difference between this conclusion and that of \citet{Gilliland2011} we ascribe to the fact that the variability metric they analyze (called CDPP) is constructed to measure the detectability of transits, and we show it is not particularly suited to assessing the sort of stellar variability under consideration here. In particular it does a poor job of sorting the stars by our much simpler and more straightforward variability measures, so that the distribution of stars in CDPP is not well-tied to the question of intrinsic stellar variability. 

We address again the question of what fraction of comparable stars in the solar neighborhood are more active (photometrically variable by our diagnostics) than the top of the comparable activity distribution for the Sun. We examine the histogram of $R_{var}(\rm{30 days})$ for the Sun over all parts of its cycle compared with our \textit{Kepler} sample to address the fact that the \textit{Kepler} stars are at unknown phases of their activity cycles. This yields the fraction of stars whose activity level at any of the random cycle phases sampled is greater than any analogous snapshot of the Sun during Cycle 23. We confirm the result of \citet{Basri2010} that this active fraction is only about one-quarter of the bright \textit{Kepler} sample. We argue that the active fraction can only reliably be determined using stars brighter than $M_{Kep}$=14 for most timescales, to avoid being misled by the contribution of noise in fainter stars. The discrepancy between our results and those of \citet{McQuillan2011}, who find a larger active fraction, is due to a combination of different choices of the level of variability that is used to characterize the active Sun, and their inclusion of stars with fainter magnitudes. Choosing the maximal activity level for the Sun (as we do) yields an active fraction of 25\%. Choosing the average level of the active Sun (as they do) yields an active fraction of 40\% on the bright \textit{Kepler} sample (but also an active fraction of 20\% for the Sun itself). The further difference that \citet{McQuillan2011} consider all stars down to 16th magnitude (where noise becomes a problem) leads to their published active fraction of 60\%, whereas we only consider solar analogs down to 12.5 magnitude. We can reproduce their results under consistent assumptions, but argue that what we have done yields the best astrophysical information. 

We also look at the active fraction for solar-type stars at timescales ranging from half an hour to a month. We find that this fraction is dominated by noise even in the bright \textit{Kepler} samples for timescales shorter than about 12 hours. For timescales of a day to a few days the active fraction as measured by MDV($t_{bin}$) is about one-third for solar-type stars. We reach the firm conclusions that a) the Sun exhibits the same levels of variability as most of the \textit{Kepler} stars on all timescales longer than half a day, and that b) the excess variability in our diagnostics of the \textit{Kepler} stars compared with the Sun at shorter timescales is not intrinsic to the stars. 

We next examine stars that are (a little) hotter and (mostly) cooler than the Sun. This is done in temperature bins with a width of 500K, starting from 6500K and ending at 3500K. We find that when one normalizes by $R_{var}$, the Sun is a fairly good predictor of MDV($t_{bin}$) for all of these stars. The exception is that late-F stars (and to a lesser extent, G stars) which have $R_{var}$ similar to that of the active Sun may in some cases be  more photometrically variable at timescales up to a few days. On the other hand (as found before), when one goes to stars cooler than the Sun the fraction of them which are more active than the active Sun increases steadily, becoming almost all of them by early M-dwarfs. Their variability at shorter timescales is also greater; stellar $R_{var}$ is a predictor of stellar MDV($t_{bin}$) at most timescales. 

We find a concentration of stars at a particular MDV over a span of brighter magnitudes for most timescales in the stars cooler than 4500K. Another way to put this is that there is a sprinkling of very quiet stars then a gap before the main levels of variability are reached for the brighter part of this cool sample. We test (but largely disconfirm) the idea that the concentration of stars at a high value of MDV could be due to contaminating giants. We are forced to reach down to fainter magnitudes as we go to cooler samples in order to have a similar sample size. This could have rendered us vulnerable to our warnings about the contribution of noise in the faintest bins, but for the cool stars the intrisic variability has grown sufficiently so that it is not a problem.

This paper develops a methodology to characterize stellar variability over very large samples of stars. It can easily be applied to other large photometric variability datasets. There are many other diagnostics that could also be applied, so this is just the beginning of a quest to get a real overview of stellar variability without examining all the individual light curves. We find that our Sun is a good model for the general photometric variability of all solar-type (convective envelope) stars, but their activity fractions depend inversely on effective temperature (assuming a fixed methodology). The next step is to continue refining general light curve structure diagnostics and use them in conjunction with $R_{var}$ and MDV to further elucidate the characteristics of the rich sample of stars in the \textit{Kepler} dataset.

\acknowledgements
This work was supported in part by NSF grant AST-0606748 for GB and a NASA \textit{Kepler} fellowship for LW. We would like to acknowledge helpful conversations with Ron Gilliland, Jon Jenkins, Jessie Christiansen, Suzanne Aigrain, and Amy McQuillan.

\end{document}